\documentclass[longauth]{aa}  

\usepackage{graphicx}
\usepackage{appendix}
\newcommand{\lya}{Ly$\alpha$~}
\begin{document}

   \title{The ALPINE-ALMA [CII] survey:}
   \subtitle{Small Ly$\alpha$-[CII] velocity offsets in main-sequence galaxies at $4.4 < z < 6$}
\titlerunning{ALPINE: small Ly$\alpha$-[CII] velocity offsets in main-sequence galaxies at $4.4 < z < 6$}

   \author{P. Cassata\inst{1,2}
    \and L. Morselli\inst{1,2}
    \and A. Faisst\inst{3}
    \and M. Ginolfi\inst{4}
    \and M. B\'ethermin\inst{5}
     \and P. Capak\inst{3,6,7}
     \and O. Le F\`evre\inst{5}
     \and D. Schaerer \inst{4,8}
     \and J. Silverman \inst{9,10}
     \and L. Yan \inst{11}
\and B. C. Lemaux\inst{12}  
\and M. Romano\inst{1,2}
\and M. Talia\inst{13,14}
\and S. Bardelli\inst{14}
\and M. Boquien\inst{15}
\and A. Cimatti\inst{13,16}
\and M. Dessauges-Zavadsky\inst{4}
\and Y. Fudamoto\inst{17,18}
\and S. Fujimoto\inst{6,7}
\and M. Giavalisco\inst{19}
\and N. P. Hathi\inst{20}
\and E. Ibar\inst{21}
\and G. Jones\inst{22,23}
\and A. M. Koekemoer\inst{20}
\and H. M\'endez-Hernandez\inst{21}
\and C. Mancini\inst{1,2}
\and P. A. Oesch\inst{4,6}
\and F. Pozzi\inst{13}
\and D. A. Riechers\inst{24,25}
\and G. Rodighiero\inst{1,2}
\and D. Vergani\inst{14}
\and G. Zamorani\inst{14}
\and E. Zucca\inst{14}
}
\authorrunning{P. Cassata et al.}

\institute{
Dipartimento di Fisica e Astronomia, Universit\`a di Padova, Vicolo dell'Osservatorio, 3 35122 Padova, Italy
\and
INAF Osservatorio Astronomico di Padova, vicolo dell'Osservatorio 5, I-35122 Padova, Italy
\and
Department of Astronomy, California Institute of Technology, 1200 E. California Blvd., MC 249-17, Pasadena, CA 91125, USA
\and
Observatoire de Gen\`eve, Universit\'e de Gen\`eve 51 Ch. des Maillettes, 1290 Versoix, Switzerland
\and
Aix Marseille Universit\'e, CNRS,  CNES, LAM (Laboratoire d'Astrophysique de Marseille), 13013, Marseille, France
\and
The Cosmic Dawn Center (DAWN), University of Copenhagen, Vibenshuset, Lyngbyvej 2, DK-2100 Copenhagen, Denmark
\and
Niels Bohr Institute, University of Copenhagen, Lyngbyvej 2, DK-2100 Copenhagen, Denmark
\and 
Institut de Recherche en Astrophysique et Plan\'etologie - IRAP, CNRS, Universit\'e de Toulouse, UPS-OMP, 14, avenue E. Belin, F31400 Toulouse, France
\and
Kavli Institute for the Physics and Mathematics of the Universe, The University of Tokyo, Kashiwa, Japan 277-8583 (Kavli IPMU, WPI)
\and
Department of Astronomy, School of Science, The University of Tokyo, 7-3-1 Hongo, Bunkyo, Tokyo 113-0033, Japan
\and 
Department of Astronomy, California Institute of Technology, 1200 E. California Blvd., MC 249-17, Pasadena, CA 91125, USA
\and 
Department  of  Physics,  University  of  California,  Davis,  One  Shields  Ave.,  Davis,  CA  95616,  USA
\and
University of Bologna, Department of Physics and Astronomy (DIFA), Via Gobetti 93/2, I-40129, Bologna, Italy
\and
INAF - Osservatorio di Astrofisica e Scienza dello Spazio di Bologna, via Gobetti 93/3, I-40129, Bologna, Italy
\and
Centro de Astronomia (CITEVA), Universidad de Antofagasta, Avenida Angamos 601, Antofagasta, Chile
\and 
INAF - Osservatorio Astrofisico di Arcetri,Largo E. Fermi 5, I-50125, Firenze, Italy,
\and
Department of Astronomy, University of Geneva, ch. des Maillettes 51, CH-1290 Versoix, Switzerland
\and
Research Institute for Science and Engineering, Waseda University, 3-4-1 Okubo, Shinjuku, Tokyo 169-8555, Japan
\and
Astronomy Department, University of Massachusetts, Amherst, MA 01003, USA
\and
Space Telescope Science Institute, 3700 San Martin Drive, Baltimore, MD 21218, USA
\and
Instituto de F\'isica y Astronom\'ia, Universidad de Valpara\'iso, Avda. Gran Breta\~na 1111, Valpara\'iso, Chile
\and
Cavendish Laboratory, University of Cambridge, 19 J. J. Thomson Ave., Cambridge CB3 0HE, UK
\and
Kavli Institute for Cosmology, University of Cambridge, Madingley Road, Cambridge CB3 0HA, UK
\and
Department of Astronomy, Cornell University, Space Sciences Building, Ithaca, NY 14853, USA 
\and 
Max-Planck Institut f\"ur Astronomie, K\"onigstuhl 17, D-69117, Heidelberg, Germany
}

   \date{Received XXX; accepted XXX}

 
  \abstract
   {The Lyman-$\alpha$ line in the ultraviolet (UV) and the [CII] line in the far-infrared (FIR) are widely used tools to identify galaxies in the early Universe and to obtain insights into interstellar medium (ISM) properties in high-redshift galaxies. By combining data obtained with ALMA in band 7 at $\sim$ 320 GHz as part of the ALMA Large Program to INvestigate [CII] at Early Times (ALPINE) with spectroscopic data from DEIMOS at the Keck Observatory, VIMOS and FORS2 at the Very Large Telescope, we assembled a unique sample of 53 main-sequence star-forming galaxies at $4.4<z<6$ in which we detect both the Lyman-$\alpha$ line in the UV and the [CII] line in the FIR.}
   {The goal of this paper is to constrain the properties of the \lya emission in these galaxies in relation to other properties of the ISM.}
   {We used [CII], observed with ALMA, as a tracer of the systemic velocity of the galaxies, and we exploited the available optical spectroscopy to obtain the \lya-[CII] and ISM-[CII] velocity offsets.}
   {We find that 90\% of the selected objects have \lya-[CII] velocity offsets in the range $0 < \Delta v_{Ly\alpha - [CII]} < $ 400 km s$^{-1}$, in line with the few measurements available so far in the early Universe, and significantly smaller than those observed at lower redshifts. At the same time, we observe ISM-[CII] offsets in the range $-500 < \Delta v_{ISM - [CII]} < $ 0 km s$^{-1}$, in line with values at all redshifts, which we interpret as evidence for outflows in these galaxies. We find significant anticorrelations between $\Delta v_{Ly\alpha-[CII]}$ and the Ly$\alpha$ rest-frame equivalent width $EW_0(Ly\alpha)$ (or equivalently, the Ly$\alpha$ escape fraction $f_{esc}(Ly\alpha)$): galaxies that show smaller $\Delta v_{Ly\alpha-[CII]}$ have larger $EW_0(Ly\alpha)$ and $f_{esc}(Ly\alpha)$.} 
{
We interpret these results in the framework of available models for the radiative transfer of \lya photons. According to the models, the escape of \lya photons would be favored in galaxies with high outflow velocities, producing large $EW_0(Ly\alpha)$ and small $\Delta v_{Ly\alpha-[CII]}$, in agreement with our observations. The uniform shell model would also predict that the \lya escape in galaxies with  slow outflows (0 < $v_{out} < 300$ km s$^{-1}$) is mainly determined by the neutral hydrogen column density (NHI) along the line of sight, while the alternative model by Steidel et al. (2010) would more highly favor a combination of NHI at the systemic velocity and covering fraction as driver of the \lya escape. We suggest that the observed increase in \lya escape that is observed in the literature between $z\sim2$ and $z\sim6$ is not due to a higher incidence of fast outflows at high redshift, but rather to a decrease in average NHI along the line of sight, or alternatively, a decrease in HI covering fraction.
   
}


   \maketitle
%

\section{Introduction}
The 1215.7\AA~Lyman-$\alpha$ line (\lya hereafter) is in principle the strongest emission line to originate from a star-forming region (Partridge \& Peebles~1967). For galaxies at $2\lesssim z \lesssim 6.5,$ it lies in the spectral regime probed by optical spectrographs on 8m class telescopes, such as the VIsible Multi-Object Spectrograph at the Very Large Telescope, VLT (VIMOS, Le F\`evre et al. 2000), the Low Resolution Imaging Spectrometer (LRIS, Oke et al. 1995), and the Deep Imaging Multi-Object Spectrograph (DEIMOS, Faber et al. 2003) at the Keck Observatory. 

For these reasons, this line has become a widely exploited tool for identifying and studying galaxies at high redshifts. On the one hand, it is used as a redshift machine, especially at $z>4$, where UV interstellar absorption features are difficult to identify over the continuum emission, which becomes very faint for reasonable integration times. It also gives important information about the physical conditions in galaxies: because it is a resonant line,  \lya is easily scattered and absorbed by neutral hydrogen HI and dust, depending on their geometry and kinematics \citep{dijkstra14}. Therefore the  \lya escape fraction  $f_{esc}(Ly\alpha$) (the ratio between the \lya flux that escapes the galaxy with respect to the flux that is produced by star formation) can be used to investigate the distribution and kinematics of the interstellar medium (ISM).

In the past 20 years, much progress has been made on the theoretical side: numerical and semianalytic simulations have predicted the emergent \lya emission from individual galaxies as a function of ISM parameters such as density, temperature, dust content, kinematics, and clumpiness, assuming spherical symmetry (Ahn~2004; Dijkstra et al. 2006; Verhamme et al. 2006; Laursen et al. 2013; Gronke \& Dijkstra 2014; Duval et al. 2014; Verhamme et al. 2015), asymmetric distributions (Zheng \& Wallace 2014, Behrens, Dijkstra \& Niemeyer 2014), and even coupling radiative transfer models to hydrodynamical simulations of realistic galaxies (Verhamme et al. 2012; Behrens \& Braun 2014). These models in general predict that ISM winds favor the escape of \lya photons, with most of the \lya emitters displaying outflows, that galaxies observed face-on in general are expected to have stronger \lya than those observed edge-on, and that a clumpier ISM, for a fixed amount of neutral hydrogen, also allows more \lya photons to escape.

From the observational point of view, its interpretation is still very complicated, although the  \lya line has been observed in galaxies at all redshifts. The Lyman Alpha Reference Sample (LARS; \"Ostlin et al. 2014) provides a unique and complete $z\sim0$ benchmark sample of different galaxies displaying diverse \lya properties, studied across the entire electromagnetic spectrum, from the UV to the 21 cm line, including H$\alpha$ imaging and spectroscopy. Despite the large amount of information available, the authors identified many factors that can affect \lya emission, however, such as dust, outflows, morphology, turbulence, and clumpiness of the ISM (Hayes 2015, Herenz et al. 2016, Guaita et al. 2015, Messa et al. 2019). A recent multivariate analysis of the low-$z$ dataset (Runnholm et al. 2020) indicates that \lya emission primarily correlates with the star formation rate (SFR; which sets the absolute scale), and then with the reddening of the stellar continuum E(B-V)$_{\ast}$ and the gas-covering fraction in decreasing order, with both governing the \lya escape.

Interestingly, an increase in average \lya escape fraction for star-forming galaxies is observed between $z\sim2$ and $z\sim6$ (Hayes et al. 2011), accompanied by an increase in the fraction of strong \lya emitters (i.e., galaxies for which the \lya line has a rest-frame equivalent width, $EW_0(Ly\alpha)$, larger than 25~\AA) over the same redshift range (Stark et al. 2010; Stark et al. 2011; Mallery et al. 2012; Cassata et al. 2015). However, the physical processes that shape this evolution are not clearly known so far (Pentericci et al. 2016), although it is tempting to interpret these results with an increase in outflow velocities, which are found to favor the escape of \lya photons (Orsi, Lacey, \& Baugh 2012; Verhamme et al. 2015). Beyond $z\sim6,$ a sudden drop in the \lya escape fraction and in the fraction of \lya emitters is observed (Hayes et al. 2011; Pentericci et al. 2011; Schenker et al. 2012; Caruana et al. 2014; Pentericci et al. 2018; Hoag et al. 2019): these effects are in general interpreted as evidence of an increasing neutral fraction of the Universe at these epochs, as expected at the end of HI reionization, with the neutral intergalactic medium (IGM) suppressing the \lya emission that escapes from galaxies. 

\begin{table*}
    \centering
\begin{tabular}{c c c c c}
\hline\hline \noalign{\smallskip} 
  ID & $\log(M_*)$ & $\log SFR$ & E(B-V)$_{\ast}$ & M$_{FUV}$\\ 
\noalign{\smallskip}
     & $M_{\odot}$ & $M_{\odot} yr^{-1}$ & \\
     \noalign{\smallskip}
          \hline
       \noalign{\smallskip}
       CG\_12    &      9.30 $^{+    0.24 }_{-    0.29}$  &    1.12 $^{+    0.20 }_{-    0.11}$  &      0.05  &     -21.31 $^{+     0.02 }_{-     0.02 }$\\
        CG\_14    &     10.49 $^{+    0.04 }_{-    0.04}$  &    1.50 $^{+    0.07 }_{-    0.07}$  &      0.05  &     -22.11 $^{+     0.03 }_{-     0.03 }$\\
        CG\_21    &      9.76 $^{+    0.19 }_{-    0.23}$  &    1.37 $^{+    0.26 }_{-    0.24}$  &      0.20  &     -20.39 $^{+     0.18 }_{-     0.13 }$\\
        CG\_38    &      9.63 $^{+    0.18 }_{-    0.20}$  &    1.54 $^{+    0.24 }_{-    0.22}$  &      0.20  &     -21.33 $^{+     0.12 }_{-     0.09 }$\\
        CG\_42    &      9.38 $^{+    0.26 }_{-    0.32}$  &    1.05 $^{+    0.26 }_{-    0.21}$  &      0.00  &     -20.60 $^{+     0.23 }_{-     0.15 }$\\
        CG\_47    &      9.09 $^{+    0.18 }_{-    0.17}$  &    1.19 $^{+    0.20 }_{-    0.16}$  &      0.10  &     -20.80 $^{+     0.13 }_{-     0.09 }$\\
     DC\_274035    &      9.42 $^{+    0.13 }_{-    0.13}$  &    1.20 $^{+    0.19 }_{-    0.09}$  &      0.05  &     -21.74 $^{+     0.18 }_{-     0.13 }$\\
     DC\_308643    &      9.77 $^{+    0.11 }_{-    0.19}$  &    1.83 $^{+    0.12 }_{-    0.22}$  &      0.10  &     -22.43 $^{+     0.08 }_{-     0.06 }$\\
     DC\_351640    &      9.67 $^{+    0.13 }_{-    0.15}$  &    1.20 $^{+    0.20 }_{-    0.12}$  &      0.00  &     -21.75 $^{+     0.62 }_{-     0.55 }$\\
     DC\_372292    &      9.44 $^{+    0.16 }_{-    0.12}$  &    1.45 $^{+    0.24 }_{-    0.25}$  &      0.10  &     -22.01 $^{+     0.30 }_{-     0.40 }$\\
     DC\_400160    &      9.65 $^{+    0.17 }_{-    0.15}$  &    1.67 $^{+    0.24 }_{-    0.23}$  &      0.10  &     -22.36 $^{+     0.09 }_{-     0.07 }$\\
     DC\_403030    &      9.68 $^{+    0.15 }_{-    0.18}$  &    1.56 $^{+    0.26 }_{-    0.28}$  &      0.15  &     -21.74 $^{+     0.17 }_{-     0.12 }$\\
     DC\_416105    &      9.45 $^{+    0.13 }_{-    0.14}$  &    1.08 $^{+    0.12 }_{-    0.11}$  &      0.00  &     -21.78 $^{+     0.61 }_{-     0.32 }$\\
     DC\_417567    &      9.81 $^{+    0.18 }_{-    0.11}$  &    1.87 $^{+    0.24 }_{-    0.23}$  &      0.10  &     -22.92 $^{+     0.11 }_{-     0.08 }$\\
     DC\_422677    &      9.85 $^{+    0.14 }_{-    0.16}$  &    1.90 $^{+    0.21 }_{-    0.19}$  &      0.20  &     -21.63 $^{+     0.15 }_{-     0.11 }$\\
     DC\_430951    &      9.45 $^{+    0.14 }_{-    0.15}$  &    1.08 $^{+    0.16 }_{-    0.11}$  &      0.00  &     -21.70 $^{+     0.83 }_{-     0.43 }$\\
     DC\_432340    &     10.10 $^{+    0.17 }_{-    0.15}$  &    1.79 $^{+    0.26 }_{-    0.26}$  &      0.15  &     -22.09 $^{+     0.16 }_{-     0.11 }$\\
     DC\_434239    &     10.32 $^{+    0.13 }_{-    0.12}$  &    2.00 $^{+    0.27 }_{-    0.28}$  &      0.25  &     -21.91 $^{+     0.15 }_{-     0.11 }$\\
     DC\_488399    &     10.20 $^{+    0.13 }_{-    0.15}$  &    1.67 $^{+    0.26 }_{-    0.27}$  &      0.05  &     -22.06 $^{+     0.40 }_{-     0.30 }$\\
     DC\_493583    &      9.61 $^{+    0.15 }_{-    0.11}$  &    1.36 $^{+    0.27 }_{-    0.22}$  &      0.10  &     -21.76 $^{+     0.15 }_{-     0.11 }$\\
     DC\_494057    &     10.15 $^{+    0.13 }_{-    0.15}$  &    1.62 $^{+    0.26 }_{-    0.20}$  &      0.00  &     -22.37 $^{+     0.33 }_{-     0.24 }$\\
     DC\_494763    &      9.53 $^{+    0.15 }_{-    0.16}$  &    1.30 $^{+    0.30 }_{-    0.30}$  &      0.15  &     -21.43 $^{+     0.33 }_{-     0.47 }$\\
     DC\_510660    &      9.47 $^{+    0.15 }_{-    0.18}$  &    1.11 $^{+    0.24 }_{-    0.11}$  &      0.10  &     -21.55 $^{+     0.22 }_{-     0.14 }$\\
     DC\_519281    &      9.87 $^{+    0.13 }_{-    0.16}$  &    1.40 $^{+    0.27 }_{-    0.21}$  &      0.00  &     -21.74 $^{+     0.30 }_{-     0.19 }$\\
     DC\_536534    &     10.36 $^{+    0.11 }_{-    0.12}$  &    1.71 $^{+    0.19 }_{-    0.27}$  &      0.05  &     -22.32 $^{+     0.32 }_{-     0.23 }$\\
     DC\_539609    &      9.38 $^{+    0.12 }_{-    0.12}$  &    1.48 $^{+    0.10 }_{-    0.08}$  &      0.00  &     -22.36 $^{+     0.34 }_{-     0.27 }$\\
     DC\_627939    &      9.98 $^{+    0.14 }_{-    0.14}$  &    1.71 $^{+    0.21 }_{-    0.26}$  &      0.10  &     -21.60 $^{+     0.18 }_{-     0.11 }$\\
     DC\_628063    &      9.86 $^{+    0.14 }_{-    0.15}$  &    1.39 $^{+    0.33 }_{-    0.27}$  &      0.20  &     -21.16 $^{+     0.29 }_{-     0.18 }$\\
     DC\_630594    &      9.77 $^{+    0.14 }_{-    0.15}$  &    1.50 $^{+    0.25 }_{-    0.29}$  &      0.20  &     -21.35 $^{+     0.27 }_{-     0.17 }$\\
     DC\_665509    &      9.75 $^{+    0.14 }_{-    0.15}$  &    1.37 $^{+    0.23 }_{-    0.15}$  &      0.05  &     -21.95 $^{+     0.16 }_{-     0.11 }$\\
     DC\_665626    &      9.21 $^{+    0.16 }_{-    0.18}$  &    0.71 $^{+    0.29 }_{-    0.18}$  &      0.00  &     -20.73 $^{+     0.33 }_{-     0.21 }$\\
     DC\_680104    &      9.23 $^{+    0.12 }_{-    0.18}$  &    1.17 $^{+    0.18 }_{-    0.11}$  &      0.05  &     -21.72 $^{+     0.17 }_{-     0.12 }$\\
     DC\_709575    &      9.68 $^{+    0.16 }_{-    0.14}$  &    1.42 $^{+    0.27 }_{-    0.28}$  &      0.15  &     -21.29 $^{+     0.20 }_{-     0.16 }$\\
     DC\_733857    &      9.54 $^{+    0.17 }_{-    0.11}$  &    1.59 $^{+    0.22 }_{-    0.21}$  &      0.10  &     -22.01 $^{+     0.13 }_{-     0.10 }$\\
     DC\_742174    &      9.56 $^{+    0.13 }_{-    0.15}$  &    1.11 $^{+    0.23 }_{-    0.13}$  &      0.00  &     -21.37 $^{+     0.35 }_{-     0.26 }$\\
     DC\_773957    &     10.00 $^{+    0.12 }_{-    0.15}$  &    1.43 $^{+    0.26 }_{-    0.24}$  &      0.00  &     -21.79 $^{+     0.27 }_{-     0.28 }$\\
     DC\_803480    &      9.22 $^{+    0.12 }_{-    0.14}$  &    1.08 $^{+    0.10 }_{-    0.08}$  &      0.00  &     -21.35 $^{+     0.25 }_{-     0.17 }$\\
     DC\_814483    &      9.91 $^{+    0.15 }_{-    0.15}$  &    1.60 $^{+    0.25 }_{-    0.23}$  &      0.10  &     -22.24 $^{+     0.12 }_{-     0.08 }$\\
     DC\_848185    &     10.37 $^{+    0.08 }_{-    0.19}$  &    2.46 $^{+    0.11 }_{-    0.21}$  &      0.20  &     -22.54 $^{+     0.37 }_{-     0.28 }$\\
     DC\_859732    &      9.77 $^{+    0.14 }_{-    0.15}$  &    1.21 $^{+    0.28 }_{-    0.26}$  &      0.05  &     -21.10 $^{+     0.37 }_{-     0.25 }$\\
     DC\_873321    &      9.97 $^{+    0.13 }_{-    0.16}$  &    1.96 $^{+    0.22 }_{-    0.17}$  &      0.15  &     -22.42 $^{+     0.38 }_{-     0.23 }$\\
     DC\_873756    &     10.25 $^{+    0.08 }_{-    0.10}$  &    0.73 $^{+    0.47 }_{-    0.22}$  &      0.00  &     -20.87 $^{+     0.32 }_{-     0.20 }$\\
     DC\_880016    &      9.75 $^{+    0.15 }_{-    0.15}$  &    1.50 $^{+    0.25 }_{-    0.29}$  &      0.20  &     -21.12 $^{+     0.27 }_{-     0.16 }$\\
     DC\_881725    &      9.96 $^{+    0.16 }_{-    0.11}$  &    1.94 $^{+    0.23 }_{-    0.29}$  &      0.25  &     -21.55 $^{+     0.22 }_{-     0.16 }$\\
   vc\_5100537582    &      9.76 $^{+    0.13 }_{-    0.15}$  &    1.17 $^{+    0.29 }_{-    0.23}$  &      0.00  &     -21.20 $^{+     0.27 }_{-     0.17 }$\\
   vc\_5100541407    &     10.12 $^{+    0.14 }_{-    0.15}$  &    1.54 $^{+    0.26 }_{-    0.23}$  &      0.10  &     -21.15 $^{+     0.34 }_{-     0.24 }$\\
   vc\_5100822662    &     10.17 $^{+    0.13 }_{-    0.14}$  &    1.82 $^{+    0.23 }_{-    0.24}$  &      0.15  &     -21.89 $^{+     0.18 }_{-     0.13 }$\\
   vc\_5100994794    &      9.73 $^{+    0.15 }_{-    0.13}$  &    1.45 $^{+    0.26 }_{-    0.30}$  &      0.15  &     -21.34 $^{+     0.30 }_{-     0.21 }$\\
   vc\_5101210235    &      9.78 $^{+    0.12 }_{-    0.15}$  &    1.41 $^{+    0.16 }_{-    0.09}$  &      0.00  &     -22.26 $^{+     0.09 }_{-     0.07 }$\\
   vc\_5101244930    &      9.67 $^{+    0.16 }_{-    0.13}$  &    1.42 $^{+    0.25 }_{-    0.23}$  &      0.10  &     -21.90 $^{+     0.16 }_{-     0.11 }$\\
    vc\_510605533    &      9.47 $^{+    0.14 }_{-    0.17}$  &    1.13 $^{+    0.16 }_{-    0.10}$  &      0.00  &     -21.48 $^{+     0.23 }_{-     0.19 }$\\
    vc\_510786441    &      9.99 $^{+    0.16 }_{-    0.17}$  &    1.60 $^{+    0.29 }_{-    0.14}$  &      0.05  &     -22.54 $^{+     0.06 }_{-     0.05 }$\\
   vc\_5180966608    &     10.82 $^{+    0.12 }_{-    0.13}$  &    2.15 $^{+    0.27 }_{-    0.25}$  &      0.20  &     -21.74 $^{+     0.17 }_{-     0.11 }$\\

\noalign{\smallskip} \hline
\noalign{\smallskip}
 
  \end{tabular}
   \caption{Properties of the sample from multiwavelength photometry. IDs starting with CG indicate objects in the CANDELS GOODS South area; IDs starting with vc indicate objects with VUDS VIMOS/VLT spectra in the COSMOS field; DC indicates objects with DEIMOS spectra in the COSMOS field.}
    \label{tab:properties_photo1}
\end{table*} 

The advent of ALMA opened a new window for the investigation of the early Universe: observations of the  [CII] 158 $\mu$m line, which enters ALMA bands 6 and 7 at $z>4$, are now standard for galaxies at these redshifts (Capak et al. 2015; Pentericci et al. 2016; Brada\v{c} et al. 2017), and [CII] has become a standard tracer for the systemic velocity (Pentericci et al. 2016; Matthee et al. 2019, 2020). [CII] is an optically thin cooling line that is not absorbed by dust and  therefore observable in the whole galaxy. It traces the cold molecular phase and photodissociation regions in star-forming galaxies (Carilli\&Walter~2013; Vallini et al. 2015; Accurso et al. 2017; Clark et al. 2019). For these reasons, it is in principle a better systemic velocity tracer than optical nebular lines (typically H$\alpha\lambda$6863~\AA, [OIII]$\lambda$5007~\AA), which are widely used as systemic velocity tracers: these tracers are heavily affected by dust attenuation, therefore they do not trace the full kinematics of galaxies, but only those of regions with moderate obscuration. 

In this paper, we investigate the kinematics and the physical properties of the ISM in a sample of 53 main-sequence star-forming galaxies at $4.4 < z < 6$, for which we exploit observations of the [CII] 158 $\mu$m line that are available as part of the ALMA Large Program to INvestigate [CII] at Early Times (ALPINE; Le F\`evre et al. 2020; B\'ethermin et al. 2020; Faisst et al. 2020), coupled with rich spectroscopic and photometric data. In particular, we analyze the offset between the \lya line and ISM lines with respect to the systemic velocity, for which [CII] can be used as proxy. We further investigate the dependence of this shift on several physical properties of the galaxies (i.e., outflow velocity, dust reddening, and \lya escape fraction) and of the \lya line itself ($EW_0(Ly\alpha)$) in order to constrain the properties of the ISM. The paper is structured as follows: in Section 2 we present the rich dataset we used, in Section 3 we analyze the \lya offset and the outflow velocity and their dependence on other physical quantities, in Section 4 we interpret our results in the framework of available models, and in Section 5 we discuss possible implications for the cosmic evolution of Lyman-$\alpha$ emitters, and we draw our conclusions.

\section{Sample and data}\label{Section:sample_data}
\subsection{Associated optical/NIR and other multiwavelength data}

\begin{table*}
    \centering
\begin{tabular}{c c c c c c c}
\hline\hline \noalign{\smallskip} 
  ID & S/N([CII]) & $EW_0(Ly\alpha)$ & $f_{esc}(Ly\alpha)$ & $z_{[CII]}$ & $\Delta_{Ly\alpha}$ & $\Delta_{ISM}$ \\
    &           & \AA & & & km s$^{-1}$ & km s$^{-1}$\\
     \noalign{\smallskip}
          \hline
       \noalign{\smallskip}
        CG\_12    &       4.4    &      30.4   $\pm     2.2  $  &     0.19 $^{+    0.05 }_{-    0.07}$  &     4.4310 $\pm     0.0005$  &     110   $\pm     85  $  &                                                         \\
        CG\_14    &       4.6    &      59.6   $\pm     3.7  $  &     0.41 $^{+    0.07 }_{-    0.06}$  &     5.5527 $\pm     0.0005$  &     334   $\pm     80  $  &                                                         \\
        CG\_21    &       4.2    &      47.7   $\pm     3.7  $  &     0.23 $^{+    0.17 }_{-    0.10}$  &     5.5716 $\pm     0.0005$  &     168   $\pm     77  $  &                                                         \\
        CG\_38    &       4.7    &      21.2   $\pm     2.1  $  &     0.05 $^{+    0.04 }_{-    0.02}$  &     5.5721 $\pm     0.0007$  &      36   $\pm     82  $  &    -127   $^{+    160   }_{-    119   }$\\
        CG\_42    &       3.7    &      53.5   $\pm     6.9  $  &     0.27 $^{+    0.16 }_{-    0.12}$  &     5.5252 $\pm     0.0005$  &     707   $\pm     36  $  &                                                         \\
        CG\_47    &       4.0    &      42.8   $\pm     5.3  $  &     0.15 $^{+    0.06 }_{-    0.05}$  &     5.5745 $\pm     0.0009$  &     378   $\pm     77  $  &                                                         \\
     DC\_274035    &       4.4    &      13.3   $\pm     1.2  $  &     0.08 $^{+    0.02 }_{-    0.03}$  &     4.4791 $\pm     0.0004$  &     202   $\pm     57  $  &    -191   $^{+     54   }_{-     38   }$\\
     DC\_308643    &       7.7    &      25.6   $\pm     1.5  $  &     0.09 $^{+    0.06 }_{-    0.02}$  &     4.5253 $\pm     0.0004$  &      32   $\pm     55  $  &    -352   $^{+    119   }_{-    119   }$\\
     DC\_351640    &       5.7    &      75.7   $\pm    27.3  $  &     0.31 $^{+    0.10 }_{-    0.11}$  &     5.7058 $\pm     0.0005$  &      75   $\pm     36  $  &                                                         \\
     DC\_372292    &       9.6    &      10.3   $\pm     1.0  $  &     0.17 $^{+    0.13 }_{-    0.07}$  &     5.1364 $\pm     0.0005$  &     -48   $\pm     44  $  &    -161   $^{+    107   }_{-     97   }$\\
     DC\_400160    &       4.5    &      10.9   $\pm     1.0  $  &     0.04 $^{+    0.03 }_{-    0.02}$  &     4.5404 $\pm     0.0007$  &     200   $\pm     58  $  &    -227   $^{+     38   }_{-     43   }$\\
     DC\_403030    &       5.0    &      10.9   $\pm     2.1  $  &     0.03 $^{+    0.02 }_{-    0.01}$  &     4.5594 $\pm     0.0004$  &     350   $\pm     36  $  &    -200   $^{+    102   }_{-     81   }$\\
     DC\_416105    &       5.3    &      71.5   $\pm    14.5  $  &     0.47 $^{+    0.14 }_{-    0.12}$  &     5.6309 $\pm     0.0005$  &     103   $\pm     40  $  &       9   $^{+     40   }_{-     45   }$\\
     DC\_417567    &       6.4    &       9.4   $\pm     1.9  $  &     0.03 $^{+    0.02 }_{-    0.01}$  &     5.6700 $\pm     0.0005$  &     386   $\pm     60  $  &    -269   $^{+     80   }_{-    107   }$\\
     DC\_422677    &       7.1    &      11.5   $\pm     1.8  $  &     0.01 $^{+    0.01 }_{-    0.00}$  &     4.4381 $\pm     0.0004$  &     308   $\pm     54  $  &    -317   $^{+    118   }_{-     63   }$\\
     DC\_430951    &       4.1    &     106.2   $\pm    27.1  $  &     0.64 $^{+    0.19 }_{-    0.20}$  &     5.6880 $\pm     0.0020$  &    -201   $\pm     63  $  &                                                         \\
     DC\_432340    &       5.5    &      17.6   $\pm     2.1  $  &     0.05 $^{+    0.04 }_{-    0.02}$  &     4.4045 $\pm     0.0004$  &      94   $\pm     39  $  &    -166   $^{+     66   }_{-     55   }$\\
     DC\_434239    &       7.4    &       4.8   $\pm     1.0  $  &     0.00 $^{+    0.00 }_{-    0.00}$  &     4.4883 $\pm     0.0004$  &     136   $\pm     64  $  &                                                         \\
     DC\_488399    &      26.2    &      18.0   $\pm     2.7  $  &     0.08 $^{+    0.07 }_{-    0.04}$  &     5.6704 $\pm     0.0005$  &     377   $\pm     60  $  &                                                         \\
     DC\_493583    &       8.3    &      24.7   $\pm     3.6  $  &     0.11 $^{+    0.08 }_{-    0.05}$  &     4.5134 $\pm     0.0004$  &     157   $\pm     45  $  &                                                         \\
     DC\_494057    &      17.1    &      22.8   $\pm     1.7  $  &     0.13 $^{+    0.08 }_{-    0.06}$  &     5.5448 $\pm     0.0005$  &      32   $\pm     42  $  &                                                         \\
     DC\_494763    &      10.5    &      47.0   $\pm    20.0  $  &     0.10 $^{+    0.09 }_{-    0.05}$  &     5.2337 $\pm     0.0004$  &     250   $\pm     60  $  &    -225   $^{+     57   }_{-     62   }$\\
     DC\_510660    &       4.0    &     100.7   $\pm    43.0  $  &     0.51 $^{+    0.14 }_{-    0.22}$  &     4.5480 $\pm     0.0010$  &     324   $\pm     59  $  &                                                         \\
     DC\_519281    &       6.7    &       7.4   $\pm     1.3  $  &     0.03 $^{+    0.02 }_{-    0.02}$  &     5.5759 $\pm     0.0005$  &     100   $\pm     53  $  &    -100   $^{+     45   }_{-     36   }$\\
     DC\_536534    &       5.0    &      24.0   $\pm     5.5  $  &     0.08 $^{+    0.07 }_{-    0.03}$  &     5.6886 $\pm     0.0006$  &     255   $\pm     85  $  &    -112   $^{+    112   }_{-     49   }$\\
     DC\_539609    &       8.9    &      29.0   $\pm     4.7  $  &     0.25 $^{+    0.05 }_{-    0.05}$  &     5.1818 $\pm     0.0005$  &      -4   $\pm     50  $  &    -385   $^{+     36   }_{-     55   }$\\
     DC\_627939    &      13.0    &       6.3   $\pm     1.8  $  &     0.01 $^{+    0.01 }_{-    0.00}$  &     4.5341 $\pm     0.0004$  &     216   $\pm     49  $  &    -303   $^{+     48   }_{-     48   }$\\
     DC\_628063    &       3.8    &      37.4   $\pm    18.6  $  &     0.09 $^{+    0.08 }_{-    0.05}$  &     4.5327 $\pm     0.0005$  &     281   $\pm     60  $  &                                                         \\
     DC\_630594    &      11.2    &      34.2   $\pm    16.8  $  &     0.09 $^{+    0.08 }_{-    0.04}$  &     4.4403 $\pm     0.0004$  &     336   $\pm     64  $  &                                                         \\
     DC\_665509    &       4.8    &      25.5   $\pm     3.8  $  &     0.20 $^{+    0.08 }_{-    0.08}$  &     4.5256 $\pm     0.0004$  &     178   $\pm     50  $  &                                                         \\
     DC\_665626    &       4.4    &      19.4   $\pm     6.9  $  &     0.12 $^{+    0.06 }_{-    0.06}$  &     4.5773 $\pm     0.0004$  &     257   $\pm     41  $  &                                                         \\
     DC\_680104    &       4.2    &      23.8   $\pm     6.3  $  &     0.16 $^{+    0.05 }_{-    0.05}$  &     4.5295 $\pm     0.0004$  &     113   $\pm     45  $  &    -439   $^{+    135   }_{-     65   }$\\
     DC\_709575    &       5.5    &      21.8   $\pm     3.4  $  &     0.12 $^{+    0.11 }_{-    0.06}$  &     4.4121 $\pm     0.0004$  &     188   $\pm     44  $  &                                                        \\
     DC\_733857    &       7.3    &      19.8   $\pm     2.3  $  &     0.08 $^{+    0.05 }_{-    0.03}$  &     4.5445 $\pm     0.0004$  &     259   $\pm     42  $  &     -32   $^{+     64   }_{-     43   }$\\
     DC\_742174    &       4.8    &      58.0   $\pm    12.6  $  &     0.20 $^{+    0.07 }_{-    0.08}$  &     5.6360 $\pm     0.0005$  &      81   $\pm     36  $  &                                                         \\
     DC\_773957    &       8.5    &      19.0   $\pm     4.4  $  &     0.10 $^{+    0.07 }_{-    0.04}$  &     5.6773 $\pm     0.0005$  &     345   $\pm     50  $  &                                                         \\
     DC\_803480    &       3.7    &      56.8   $\pm    14.7  $  &     0.28 $^{+    0.05 }_{-    0.06}$  &     4.5417 $\pm     0.0004$  &     -27   $\pm     38  $  &                                                         \\
     DC\_814483    &       4.6    &       5.2   $\pm     1.0  $  &     0.04 $^{+    0.03 }_{-    0.02}$  &     4.5810 $\pm     0.0004$  &     155   $\pm     50  $  &    -201   $^{+    126   }_{-     89   }$\\
     DC\_848185    &      18.3    &       5.3   $\pm     1.0  $  &     0.00 $^{+    0.00 }_{-    0.00}$  &     5.2931 $\pm     0.0004$  &     509   $\pm     48  $  &    -292   $^{+     93   }_{-     69   }$\\
     DC\_859732    &       4.3    &      21.0   $\pm    10.6  $  &     0.19 $^{+    0.16 }_{-    0.09}$  &     4.5318 $\pm     0.0005$  &     135   $\pm     57  $  &    -276   $^{+     97   }_{-    108   }$\\
     DC\_873321    &       7.5    &       8.3   $\pm     1.0  $  &     0.08 $^{+    0.04 }_{-    0.03}$  &     5.1542 $\pm     0.0004$  &     180   $\pm     45  $  &    -565   $^{+    107   }_{-    346   }$\\
     DC\_873756    &      32.6    &      19.9   $\pm     3.6  $  &     0.32 $^{+    0.21 }_{-    0.21}$  &     4.5457 $\pm     0.0004$  &     129   $\pm     71  $  &     199   $^{+     32   }_{-     37   }$\\
     DC\_880016    &       8.6    &       5.9   $\pm     1.0  $  &     0.01 $^{+    0.01 }_{-    0.00}$  &     4.5415 $\pm     0.0004$  &     254   $\pm     53  $  &    -470   $^{+    183   }_{-    129   }$\\
     DC\_881725    &      12.3    &      10.3   $\pm     3.2  $  &     0.01 $^{+    0.01 }_{-    0.00}$  &     4.5777 $\pm     0.0004$  &     408   $\pm     46  $  &    -247   $^{+    177   }_{-     80   }$\\
   vc\_5100537582    &       8.1    &      19.3   $\pm     1.2  $  &     0.09 $^{+    0.06 }_{-    0.04}$  &     4.5501 $\pm     0.0004$  &      -4   $\pm    146  $  &    -377   $^{+    178   }_{-    162   }$\\
   vc\_5100541407    &      11.4    &      15.7   $\pm     2.4  $  &     0.02 $^{+    0.02 }_{-    0.01}$  &     4.5630 $\pm     0.0004$  &      11   $\pm    146  $  &                                                         \\
   vc\_5100822662    &      14.9    &       7.6   $\pm     1.0  $  &     0.01 $^{+    0.01 }_{-    0.00}$  &     4.5205 $\pm     0.0004$  &     403   $\pm    242  $  &    -255   $^{+    119   }_{-    152   }$\\
   vc\_5100994794    &      12.0    &       9.9   $\pm     1.2  $  &     0.02 $^{+    0.02 }_{-    0.01}$  &     4.5802 $\pm     0.0004$  &     268   $\pm    192  $  &                                                         \\
   vc\_5101210235    &       4.3    &      25.4   $\pm     1.0  $  &     0.20 $^{+    0.05 }_{-    0.06}$  &     4.5761 $\pm     0.0004$  &      16   $\pm    144  $  &    -182   $^{+    177   }_{-    139   }$\\
   vc\_5101244930    &       5.0    &      22.4   $\pm     1.4  $  &     0.11 $^{+    0.08 }_{-    0.05}$  &     4.5803 $\pm     0.0006$  &      27   $\pm    150  $  &    -225   $^{+     42   }_{-    112   }$\\
    vc\_510605533    &       4.9    &      12.8   $\pm     1.0  $  &     0.06 $^{+    0.02 }_{-    0.02}$  &     4.5019 $\pm     0.0006$  &     460   $\pm    146  $  &                                                         \\
    vc\_510786441    &      11.1    &      32.4   $\pm     1.4  $  &     0.25 $^{+    0.09 }_{-    0.12}$  &     4.4635 $\pm     0.0004$  &     157   $\pm    197  $  &                                                         \\
   vc\_5180966608    &      12.5    &     117.9   $\pm    19.1  $  &     0.02 $^{+    0.02 }_{-    0.01}$  &     4.5296 $\pm     0.0004$  &     -90   $\pm    195  $  &                                                         \\

   \noalign{\smallskip} \hline
\noalign{\smallskip}
 
  \end{tabular}
   \caption{\lya, [CII], and ISM measurements for the sample. The IDs are the same as in Table~\ref{tab:properties_photo1}.}
    \label{tab:properties_photo}
\end{table*} 

The galaxies for this study have been selected from ALPINE (Le F\`evre et al. 2020; B\'ethermin et al. 2020; Faisst et al. 2020), an ALMA project with the main goal of revealing [CII], and the underlying $\sim158~\mu$m rest-frame continuum, in a sample of 118 main-sequence galaxies at $4.4 < z < 6$ in the COSMOS (Scoville et al. 2007) and GOODS-S fields (Giavalisco et al. 2004). The sample is originally drawn from large spectroscopic surveys of normal star-forming galaxies: the bulk of the spectroscopic redshifts come from extensive campaigns at the VLT (VUDS, Le F\`evre et al. 2015; Vanzella et al. 2008; Balestra et al. 2010) and Keck (DEIMOS 10k survey, Hasinger et al. 2018). The availability of high-quality spectroscopic redshifts is crucial to ensure that the expected [CII] emission is redshifted in a high transmission atmospheric window within ALMA band 7. This restricts the sample to two subranges in redshift, $4.4 < z <4.65$ and $5.05 < z < 5.8$. More details can be found in Le F\`evre et al. (2020).

A wealth of ancillary data is available for the targets in the COSMOS field as part of the many observational projects, including the i band of the Hubble Space Telescope (HST) (Scoville et al. 2007, Koekemoer et al. 2007), Subaru optical imaging (Taniguchi et al. 2007), deep NIR YJHK imaging from the UltraVista Survey (McCracken et al. 2012), Spitzer 3.6 and 4.5 $\mu$m imaging (Sanders et al. 2007; Laigle et al. 2016), with good X-ray coverage with both XMM-Newton (Hasinger et al. 2007) and Chandra (Elvis et al. 2009; Civano et al. 2016), as well as radio waves (Smol{\v{c}}i{\'c} et al. 2017). A similar wealth of data is also available in the GOODS-S field, including HST optical photometry with HST from the GOODS project (Giavalisco et al. 2004), HST/NIR photometry with the WFC3 camera from the CANDELS survey (Grogin et al. 2011, Koekemoer et al. 2011), $K_s$ photometry with HAWK-I/VLT from HUGS (Fontana et al. 2014),  NIR photometry between 3.6 and 8 $\mu$m with Spitzer, and with deep and wide X-ray coverage with Chandra (Luo et al. 2017).

These observations allow us to build spectral energy distributions (SEDs) from the UV to the NIR rest-frame that we fit with standard procedures to obtain key quantities such as stellar mass, star-formation rate, absolute magnitude in the far-UV M$_{FUV}$, and E(B-V)$_{\ast}$. The SED fitting parameters that are used in this paper are presented in Table~\ref{tab:properties_photo1}, along with their uncertainties; for more details, see Faisst et al. (2020), where the ancillary data for the ALPINE galaxies are presented in their full extent. The ALPINE galaxies, as expected, are mainly main-sequence galaxies with stellar masses $3\times10^{8} < M/M_{\odot} < 10^{11}$, SFRs $3 < SFR < 300 M_{\odot} yr^{-1}$, and far-UV absolute magnitudes in the range $-23 < M_{FUV} < -20.5$, which align perfectly with the main-sequence parameterization for $z\sim5$ by Speagle et al. (2014). Most importantly, all the samples that we used as comparison at lower redshift (Shapley et al. 2003, Steidel et al. 2010, Talia et al. 2017, Marchi et al. 2019) also mainly include main-sequence galaxies that largely overlap in stellar mass with our sample, with differences in SFR and $M_{FUV}$ that are compatible with the evolving main-sequence from  $z\sim2$ to $z\sim5$. In particular, at $z\sim2.3,$ the sample in Steidel et al. (2010) covers masses $10^9<M^*<10^{11} M_{\odot}$, $10<SFR<100 M_{\odot} yr^{-1}$, and far-UV absolute magnitudes $-22.5 < M_{FUV} < -19.7$ (see Erb et al. 2006a,b), with ranges that are very similar to those reported by Du et al. (2018) and Pahl et al. (2020). The 800 galaxies at $z\sim3$ that are stacked in Shapley et al. (2003) have $-22.6 < M_{FUV} < -20.3$, with an average $M_{FUV} \sim -21.4$, and have SFRs $25<SFR<50 M_{\odot} yr^{-1}$.   Marchi et al. (2019) have $3x10^8<M^*/M_{\odot}<10^{11}$ and $3  < SFR < 100 M_{\odot} yr^{-1}$, which is very similar to our sample.

\subsection{ALMA observations and reduction}
The ALMA observations were carried out in band 7 starting in May 2018 during cycle 5 and completed in February 2019 in cycle 6, in configurations C43-1 and C43-2, which ideally produce angular resolutions coarser than 0.7 arcsec (in order not to over-resolve the flux for targets that have expected sizes of 0.5-0.7 arcsec, Capak et al. 2015). Integration times range between 15 and 30 minutes, and the spectral setting for each object was adjusted to set the spectral windows around the expected frequency of [CII] emission, based on the UV spectroscopic redshift. The final spectral resolution varies with redshift from 26 to 35 km s$^{-1}$ and the median beam size of the ALMA observations is about 1.13" x 0.85" full width at half maximum (FWHM). For more details, see B\'ethermin et al. (2020), where the processing of ALMA data is presented in full detail.

The data revealed bright [CII] flux (S/N $>$ 3.5) for 75 galaxies (see Le F\`evre et al. 2020, B\'ethermin et al. 2020 for details). For these galaxies, we can derive reliable systemic velocities, as traced by the [CII] emission. In particular, when the ALMA spectrum is extracted for the [CII] emitting region, a Gaussian is fit to the spectrum and the centroid of the Gaussian is used to determine $\nu_{[CII]obs}$, $z_{[CII]obs}$, and in this way, the systemic velocity of the system (B\'ethermin et al. 2020). The typical error on $\nu_{[CII]obs}$ depends on the width of the [CII] line and corresponds to an error on $z_{[CII]obs}$ of 0.0005-0.0010 (see Table~\ref{tab:properties_photo}). As expected, $z_{[CII]obs}$ is in general different than the spectroscopic redshift $z_{spec}$, which has been measured by combining different UV rest-frame tracers (mainly \lya and the UV rest-frame ISM lines).

\subsection{Spectroscopic data and measurements}
The full details of the spectroscopic data can be found in Faisst et al. (2020). Here we briefly present the data analysis that is relevant for this paper.

Out of the 75 galaxies for which [CII] is detected at a signal-to-noise ratio (S/N) $>3.5$, 8 are in the GOODS-S field: 7 have spectra obtained with VIMOS and FORS2 at the VLT (Vanzella et al. 2008; Balestra et al. 2010), with spectral resolutions $R\sim600,$ and the remaining galaxy has a spectrum obtained from the HST/GRAPES grism data (Malhotra et al. 2005). The remaining 66 galaxies lie in the COSMOS field: 49 have spectra from DEIMOS at Keck (Capak et al. 2004; Mallery et al. 2012; Hasinger et al. 2018), with a spectral resolution R $\sim2500$, and 17 have spectra from VIMOS (VUDS survey; Le F\`evre et al. 2015), with a lower spectral resolution R $\sim230$. In order to reduce the noise of the DEIMOS spectra, which also affects the shape of the \lya line, we applied a boxcar smoothing to those spectra, with a window length of 100 km s$^{-1}$.

We computed the $EW_0(Ly\alpha)$, defined as the ratio of the flux of the \lya emission and the flux density of the underlying continuum, from the calibrated spectra to the photometry (see Faisst et al. 2020 for details). We used the continuum redward of the $Ly\alpha$ line ($\lambda_{Ly\alpha}$), as we expect the continuum at $\lambda < \lambda_{Ly\alpha}$ to be affected by ISM absorption. We estimated the continuum as the average flux value in the wavelength window [1225:1245]\AA~(an amplitude of $\sim$ 120\AA~in the observed spectra), thus considering a region without strong absorption lines. Then, we moved the window of 1\AA~towards larger wavelengths (i.e., [1226:1246]\AA) and obtained a second value of the continuum. We repeated this procedure for a total of ten windows and took the final value of the continuum for the EW calculation as the median of the ten estimates. The error on the continuum, which we consider as the main source of error on the EW estimate, was taken as the error on the mean in the first interval (i.e., the standard deviation of the flux divided by the square roots of the elements in the [1225:1245]\AA~range). The typical error on the continuum is $\sim 15\%$. All the spectra were visually inspected to confirm that no strong features and corrupted wavelengths enter the wavelength range where the continuum was computed. Finally, we computed the total $Ly\alpha$ flux ($F_{Ly\alpha}$) by summing the flux values over the wavelengths in the \lya region for which the flux is higher than the continuum. 

The \lya emission is strong enough ($EW_0$(Ly$\alpha)>4$~\AA)  for 53 of the 75 galaxies for which [CII] is detected at S/N$>3.5$  to reliably measure the velocity offset between \lya and [CII]. Four galaxies were discarded because of different problems\footnote{in two cases, the velocity offset between \lya and [CII] exceeds 2000 kms$^{-1}$,  likely because \lya and [CII] originate from two different galaxies; in one case, the [CII] emission from the object is contaminated by a nearby companion; and in the remaining case, the multiwavelength photometry is too noisy and does not allow robustly estimating stellar mass, SFR, and \lya escape fraction.} , and  in the remaining 18,  \lya is either too weak or in absorption.

In order to allow for meaningful comparisons with galaxies at lower redshift, it is important to establish if our sample is in some way biased towards strong \lya emitters. The classical definition of \lya emitter (LAE) is an object with $EW_0>20$~\AA (Ouchi et al. 2003), but the authors who establish the increase in the fraction of strong LAE with redshift (Stark et al. 2011, Mallery et al. 2012, Cassata et al. 2015) use a more stringent definition with $EW_0>25$~\AA. We checked how many galaxies in our sample have $EW_0(Ly\alpha)$ in excess of those two values. We find that, among the 53 galaxies with detected $Ly\alpha$, 20(28) have $EW_0(Ly\alpha)>25(20)$~\AA: this implies that the number of LAEs among the sample of [CII] detected galaxies is 20(28) out of 71\footnote{We excluded the 4 objects that we discarded because of various problems from the 75 detections.}, corresponding to fractions of 28(39)\%. We verified that this fraction is very close to the fraction we obtain for the whole ALPINE target sample: 39(51) out of the 117 targets\footnote{Here we excluded only the object for which the [CII] emission is contaminated by a bright companion.} have $EW_0(Ly\alpha)>25(20)$~\AA, corresponding to 33(44)\%. This implies that the probability of detecting [CII] is not strongly dependent on the presence of bright \lya emission (see Schaerer et al. 2020 for more details). These fractions are slightly larger than those found in the literature for populations of typical star-forming galaxies at similar redshifts: Stark et al. (2010), Stark et al. (2011), Mallery et al. (2012), and Cassata et al. (2015) all reported fractions of LAEs with $EW_0(Ly\alpha)>25$~\AA~ of about 25\% at $z\sim4-5$. We verified that the slightly higher fraction of strong emitters in ALPINE arises because when we selected targets, we preferentially excluded galaxies with weak \lya emission ($0<EW_{Ly\alpha}<4$) and no continuum, for which it is harder to obtain a robust and reliable spectroscopic redshift. A redshift with these qualities is crucial to  ensure that the expected [CII] emission falls within the $\sim6000$ km s$^{-1}$ ALMA spectral window. As a result, we have a slightly smaller global population, which in turn results in a slightly higher LAE fraction. This implies at the same time that the our selection is not biased toward strong \lya emitters and that we select a sample of typical galaxies with \lya in emission.

For each of the 53 galaxies we measured the velocity offset between \lya and the systemic redshift as traced by [CII]. We described in Section~\ref{Section:sample_data} that the systemic redshift is estimated as the centroid of the Gaussian fit to the [CII] emission. For \lya, we used the peak of the emission to determine $z_{Ly\alpha}$ and the velocity difference. The choice of using the peak and not a fit to the line is justified by two facts: 1. the \lya line is asymmetric, with the blue tail absorbed by ISM, and fitting a half Gaussian to the red part of the line does not give reliable and stable results; 2. most of the theoretical works present their predictions for the \lya line morphology using the peak of the line to determine the velocity offsets (e.g., Verhamme et al. 2008; Marchi et al. 2019). 

In order to estimate the random and systematic errors on the measure of the \lya position with data with very different spectral resolutions, we ran some simulations and tests. Starting with the data with the best spectral resolution, that is, galaxies with DEIMOS spectroscopy, we first verified that the smoothing we applied to reduce the noise (boxcar with window length 100 km s$^{-1}$) increases the width of the line only marginally, and more importantly, that it does not affect the position of the peak of the line. By simulating asymmetric lines at the same spectral resolution as the DEIMOS spectra, we find that the error associated with the position of the peak is about 1/5 of the width of the line; therefore, the typical error on the position of \lya for DEIMOS spectra is about 25 km s$^{-1}$ (depending on the width of each line). For galaxies in the CANDELS field (with spectral resolutions R$\sim$600) and galaxies coming from the VUDS survey (with spectral resolutions R$\sim$230), we verified again that the random error in the positioning of the line is about 1/5 of the width of the line: this typically corresponds to errors of 70 km s$^{-1}$ for CANDELS galaxies and of 150 km s$^{-1}$ for VUDS. We propagated the uncertainties on $z_{[CII]}$ (15-50 km s$^{-1}$ depending on the width of the line) and $z_{Ly\alpha}$ on the velocity offset $\Delta v_{Ly\alpha}$, and obtained errors on the \lya velocity offsets of $\sim$50, $\sim$80, and $\sim$150km/s for galaxies with DEIMOS, CANDELS, and VUDS spectroscopy, respectively (see Table~\ref{tab:properties_photo1}).

Moreover, in order to quantify the possible additional reddening of the \lya peak due to the resolution effects (the peak of an intrinsically narrow but asymmetric \lya line would move redward if it were convoluted with a wide Gaussian) we ran some simple experiments: we degraded the DEIMOS spectra at the resolution of CANDELS and VUDS, and we measured the shift of the \lya peak induced by degrading the resolution. As the \lya lines in the DEIMOS spectra have different spectral shapes (as a result of different combinations of width and skewness), we obtain a distribution of additional reddenings when we degrade them: the more symmetric the line, the lower the additional reddening. We find that for CANDELS, the distribution of the additional reddenings ranges between +20 and +80 km s$^{-1}$, with a peak at 50 km s$^{-1}$, and for VUDS, the distribution ranges between +50 and +220 km s$^{-1}$, with a peak at 150 km s$^{-1}$. Therefore we applied an average correction of 150 and 50 km s$^{-1}$ to the offsets between \lya and [CII] for the VUDS and CANDELS galaxies, respectively.

The redshift of our sample is high enough ($z>4.5$) that intergalactic absorption from neutral hydrogen might play a significant role in shaping the \lya line profile. This would significantly reduce the blue tail (Laursen, Sommer-Larsen \& Razoumov 2011). However, because of the large cosmic variance associated with this effect (the HI in the Universe is very clumpy, and therefore some lines of sight can be completely free of HI while others may be severely affected, see Laursen, Sommer-Larsen \& Razoumov 2011), we preferred not to correct the individual line profiles for this effect. However, we note that only in galaxies with the smallest $\Delta v_{Ly\alpha}$ might the red peak be appreciably reduced (and possibly shifted by the ISM absorption to redder wavelengths): in these 20\% of the cases, it is possible in principle that we underestimate the \lya flux and overestimate the $\Delta v_{Ly\alpha}$ (although it is unlikely that this is a systematic effect: these galaxies are already the galaxies with the largest $f_{esc}$(Ly$\alpha)$ and the smallest $\Delta v_{Ly\alpha}$). Pahl et al. (2020) have shown that the red peak in a composite of $\sim$200 $z\sim5$ galaxies is only marginally affected by the IGM absorption.

We also estimated the velocity offsets between the UV rest-frame ISM lines and the systemic velocity. The details of how $z_{ISM}$ is determined are given in Faisst et al. (2020): to summarize, we used an automatic tool to cross-match the Shapley~et~al.~(2003) template with our spectra, identifying features as SiII$\lambda$1260.4~\AA, OI+SiII$\lambda$1303~\AA, [CII]$\lambda$1334.5~\AA, SiIV$\lambda$1393.8+SiIV$\lambda$1402.8~\AA, and SiII$\lambda$1526.7+CIV$\lambda$1548.2~\AA+CIV$\lambda$1550.8~\AA.  When it is seen in emission, \lya was not used in the cross-correlation process. The errors were estimated by repeating the cross-match 200 times, each time perturbing the fluxes according to a Gaussian error distribution, with $\sigma$ defined by the average flux noise of the continuum (Faisst et al. 2020), and the errors are about 80-100 km s$^{-1}$. These measurements are more complex than for Ly$\alpha$ because different ISM lines have far smaller EWs than \lya, are detected with different significance for each galaxy, and some of them might overlap, depending on the redshift, with skylines or noisy parts of the spectra. We therefore visually inspected all the fits to exclude peak noises that are misinterpreted as true lines, and we assigned a reliability flag that depends on how many good lines are fit for each spectrum. We used only objects for which at least two clean features were identified and that passed a visual quality assessment in our analysis. 

Another important parameter that we need for our analysis is the \lya escape fraction ($f_{esc}$(Ly$\alpha)$): it is defined as the ratio between the flux of \lya photons that can escape the galaxy and reach the observer, and the intrinsic flux of \lya photons that is produced by star formation. The observed \lya flux is the same as we used to estimate the \lya EW, while the intrinsic flux is obtained starting from the observed SFR that we obtained from the SED fitting (Faisst et al. 2020), and converting it into intrinsic flux using the conversion by Kennicutt~(1998), which is valid for case B recombination (Brocklehurst 1971) and solar metallicity, and taking into account the appropriate conversion between a Salpeter (assumed in Kennicutt~1998) and Chabrier IMF (used here). Because galaxies at high redshift typically have subsolar metallicities (Troncoso et al. 2014; Onodera et al. 2016), we also corrected the intrinsic \lya fluxes that we obtained from the Kennicutt relation for the following effect: when we assume a metallicity of one-fifth solar (the typical value for main-sequence galaxies found by Faisst et al. 2016 at $z\sim5$), the inferred \lya fluxes are expected to be 0.2 dex higher for the same SFR (Ly et al. 2016). The error on the $f_{esc}$(Ly$\alpha$) is obtained by propagating the error on the \lya flux and on the SFR; the error on the SFR dominates the error budget. 

All the relevant measurements we used are reported in Table~\ref{tab:properties_photo}, along with their uncertainties. Objects for which  $z_{ISM}$ was considered unreliable (estimated from fewer than two lines, or from spectra with low S/N) do not have $z_{ISM}$ in this catalog.

 %
   \begin{figure}[!ht]
   \centering
   \includegraphics[width=\columnwidth]{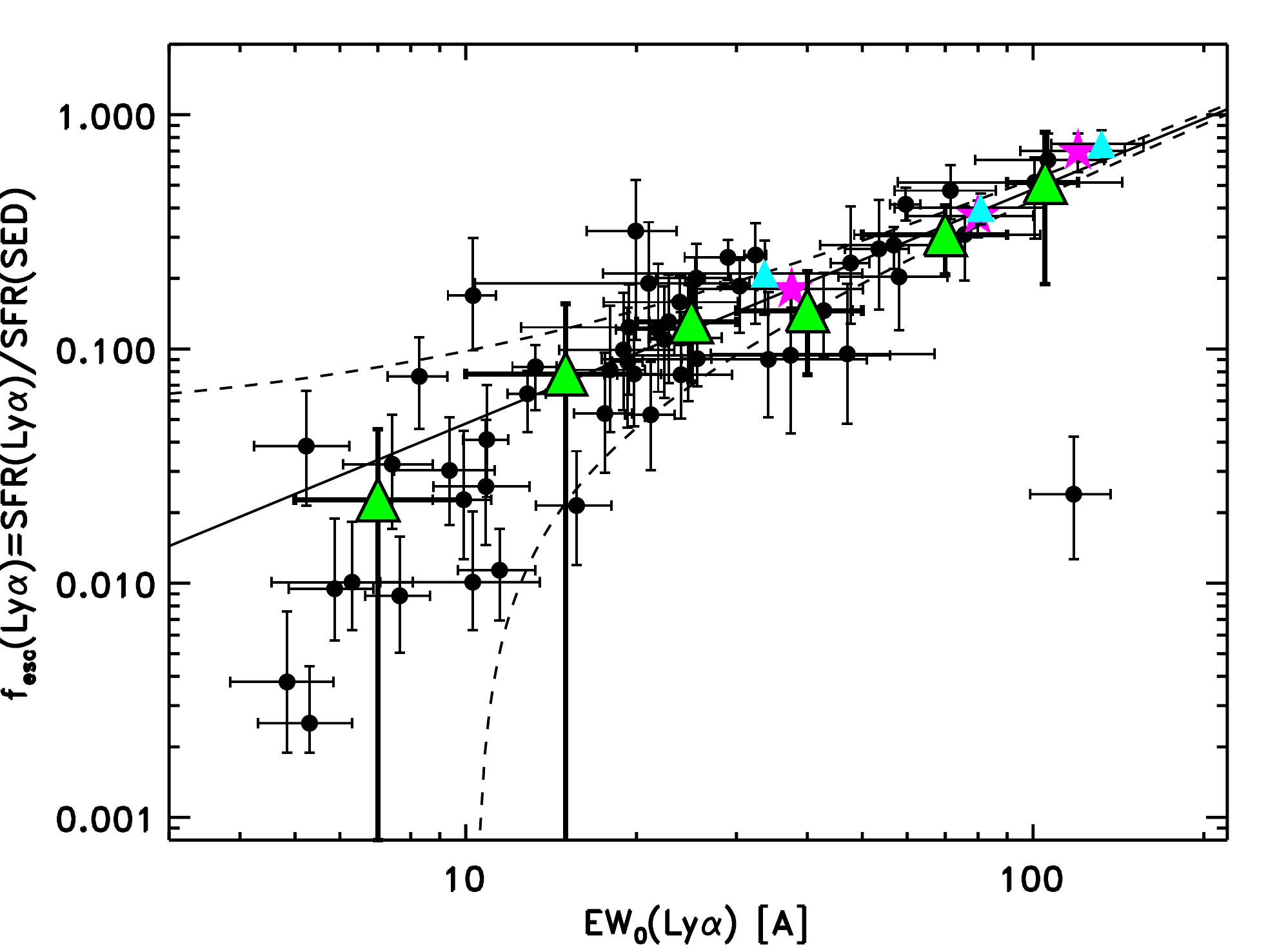}
   \caption{$f_{esc}(Ly\alpha$) as a function of $EW_0$(Ly$\alpha)$
    for the galaxies in this sample. Green triangles show $f_{esc}(Ly\alpha$) in bins of $EW_0(Ly\alpha)$. Cyan triangles and magenta stars are the average values for samples at $z\sim2.2$ and $z\sim2.6$ by Sobral et al. (2017) and Trainor et al. (2016), respectively, and the black solid and dashed lines indicate the empirical fit $f_{esc}(Ly\alpha) = 0.0048\times EW_0 \pm 0.05$ proposed by Sobral~\&~Matthee (2019).
        }
      \label{Fig:escape_vs_ew}
   \end{figure}

Figure~\ref{Fig:escape_vs_ew} shows $f_{esc}$(Ly$\alpha)$, as a function of $EW_0$(Ly$\alpha)$: the two quantities correlate well, and our measurements are spread around the empirical correlation proposed by Sobral~\&~Matthee (2019) with only a few outliers. Although most galaxies with $EW_0$(Ly$\alpha)<10$~\AA~  deviate below the relation, 65\% of the objects are within $\pm1\sigma$ from it. The average values agree well with those determined at $z\sim2.2$ and $z\sim2.9$ by Sobral et al. (2017) and Trainor et al. (2016), respectively. A correlation between these two quantities is expected: a high escape fraction means that for the same amount of star formation, or equivalently, UV photons, more \lya photons escape from the galaxy; therefore the equivalent width, that is, the ratio between \lya flux and the underlying UV continuum, increases as well. This correlation also reassures us that the \lya escape fractions that we derived by comparing the \lya flux to the SFR derived from the SED fitting are reasonable. 

\subsection{Spatial coherence between [CII] and Ly$\alpha$ emitting regions}
As presented in Le F\`evre et al. (2020), some of the ALPINE galaxies may appear very different in the UV-optical rest-frame and in the [CII] maps: in particular, although only one component might appear in the [CII] map, some objects are composed of many different bright clumps in the UV-optical. According to the morpho-kinematic classification presented in Le F\`evre et al. (2020), $\sim$40\% of ALPINE galaxies are mergers, $\sim$30\% are dispersion-dominated objects, $\sim$15\% are rotators, and the rest are too faint to be classified. 

It is therefore important to discuss how coherent the regions are that emit [CII] and Ly$\alpha$ for the galaxies we studied. We provide this discussion in the appendix, where we show for each object the [CII] map, overlaid with the contours of UV-optical images, and the position of the spectroscopic slit. In this way, we are able to compare the positions at which the \lya flux is extracted with those for which the [CII] line is used to trace the systemic redshift of the system. 

Summarizing the result from the appendix, to which we refer for all the details, the main conclusion is that \lya originates from the same object that emits the [CII] line. In some cases, we find spatial offsets between \lya and [CII] of about 0.5", which can be explained with an inhomogeneous gas and/or dust distribution in these objects. For the many mergers that are in the sample, the analysis presented here applies to the main component of the merger (that emits the bulk of the [CII] and of the \lya flux).

\section{Results}
\subsection{ISM kinematics inferred from \lya and ISM velocity offsets}

\begin{figure}[!ht]
   \centering
   \includegraphics[width=\columnwidth]{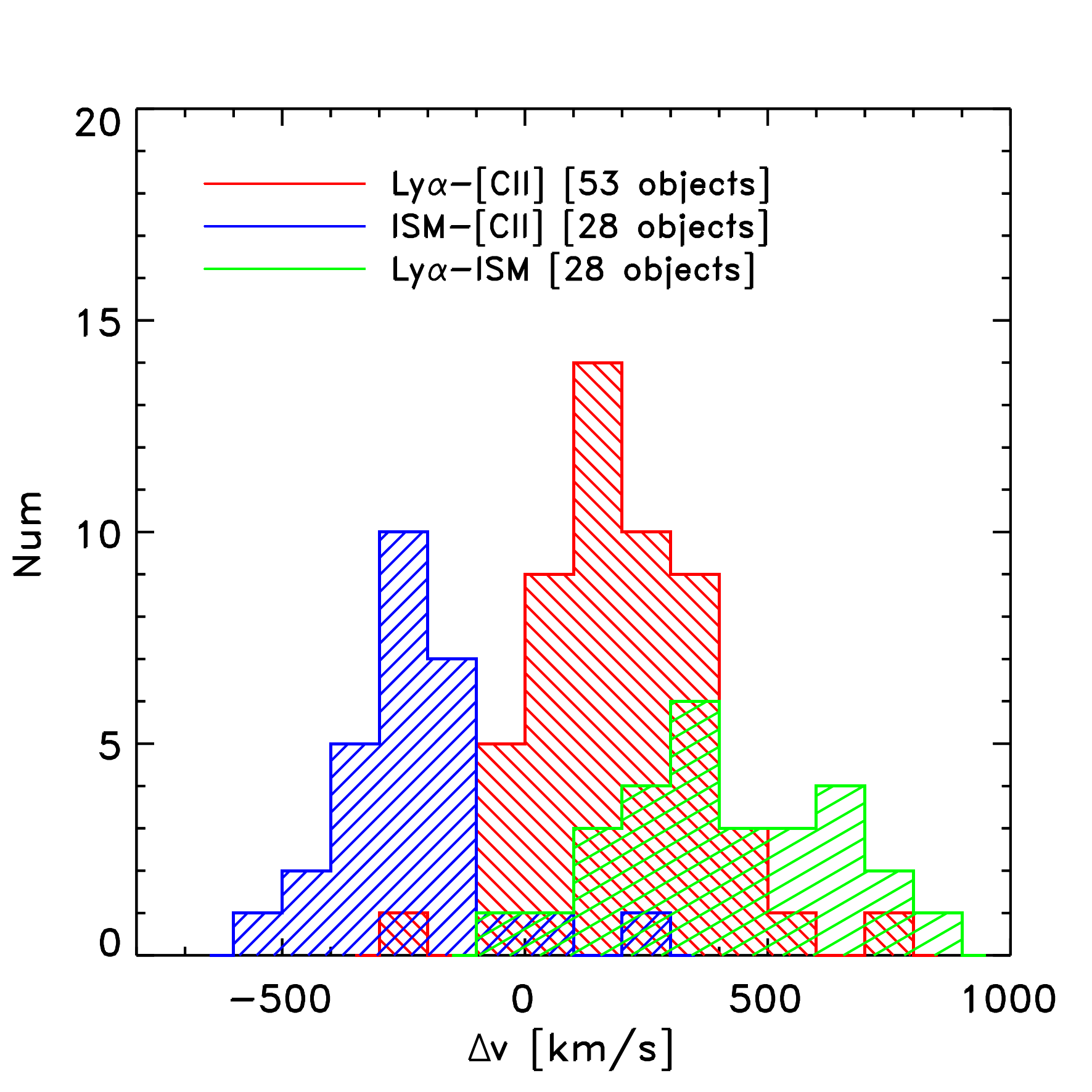}
   \caption{Red histogram: Offset velocities between \lya and [CII] for the 53 galaxies in this work; the blue histogram shows the offset velocities between ISM lines and [CII] for the 28 of the original 53 objects with \lya for which we were also able to reliably estimate $z_{ISM}$, using at least two ISM lines; the green histogram shows  the velocity difference between \lya and ISM lines for the same 28 galaxies.
        }
      \label{Fig:hist_delta}
   \end{figure}

Figure~\ref{Fig:hist_delta} shows the histograms of the velocity offsets between \lya and the ISM lines with respect to the systemic velocity, traced by [CII], and the velocity difference between \lya and ISM lines. It is clear that \lya is almost always redshifted with respect to [CII], with velocity offsets peaking at $\sim$200 km s$^{-1}$ and ranging from -200 to +600 km s$^{-1}$. The \lya offsets found in this work are consistent with the few available measurements at $z>5$ (Pentericci et al. 2016; Brada\v{c} et al. 2017; Matthee et al. 2019, 2020), who also found  offsets between \lya and [CII] around +200 km s$^{-1}$ for a handful of galaxies. However, the offsets are smaller on average than those found at $z\sim2$ and $z\sim3$ by Erb et al. (2004) and Steidel et al. (2010), and also smaller than the recent estimates at $z\sim3.5$ by Marchi et al. (2019), who all found velocity offsets of  $\sim 400-500$ km s$^{-1}$. 

In the same figure we also show the offset $\Delta v_{ISM}$ between the ISM lines and the systemic velocity traced by [CII] for the subsample of 28 objects with a robust determination of $z_{ISM}$ (these objects have at least two well-detected ISM lines): as found in the literature, the ISM lines are almost always blueshifted with respect to the systemic velocity, with offsets around -300 km s$^{-1}$ and ranging from -500 and 100 km s$^{-1}$. These offsets are similar to the values found in the literature at lower redshifts:  Erb et al. (2004) and Steidel et al. (2010) reported offsets between -400 and 100 km s$^{-1}$ at $z\sim2.3$, with an average of -150 km s$^{-1}$; Talia et al. (2017) reported an average offset of -150 km s$^{-1}$  at $z\sim2.3;$  and Marchi et al. (2019) found offsets between -400 and 100 km s$^{-1}$ with an average of $\sim$-100 km s$^{-1}$  at $z\sim3.5.$ 

In the same figure we also report for the subsample of 28 objects with a robust determination of $z_{ISM}$ the offset between \lya and the ISM lines: we find a broad distribution for this velocity difference, with an average value of 377, a scatter of 329 km s$^{-1}$, and a standard deviation of the mean $\sigma_{mean}$ of 62 km s$^{-1}$. This distribution largely overlaps the distributions for samples at similar redshifts by Pahl et al. (2020) and Faisst et al. (2016), who found average offsets of 496 ($\sigma=$ 222 km s$^{-1}$, $\sigma_{mean}=$ 26 km s$^{-1}$) and 429 ($\sigma$=229 km s$^{-1}$, $\sigma_{mean}=$ 15 km s$^{-1}$), respectively, at $z\sim5$ (although the averages agree to within 1 $\sigma_{mean}$ only between our sample and the sample of Faisst et al. 2016). At $z\sim2-3,$ Shapley et al. (2003) and Steidel et al. (2010) find an average \lya - ISM velocity difference of $\sim$ 600 km s$^{-1}$.

   \begin{figure}[!ht]
   \centering
   \includegraphics[width=\columnwidth]{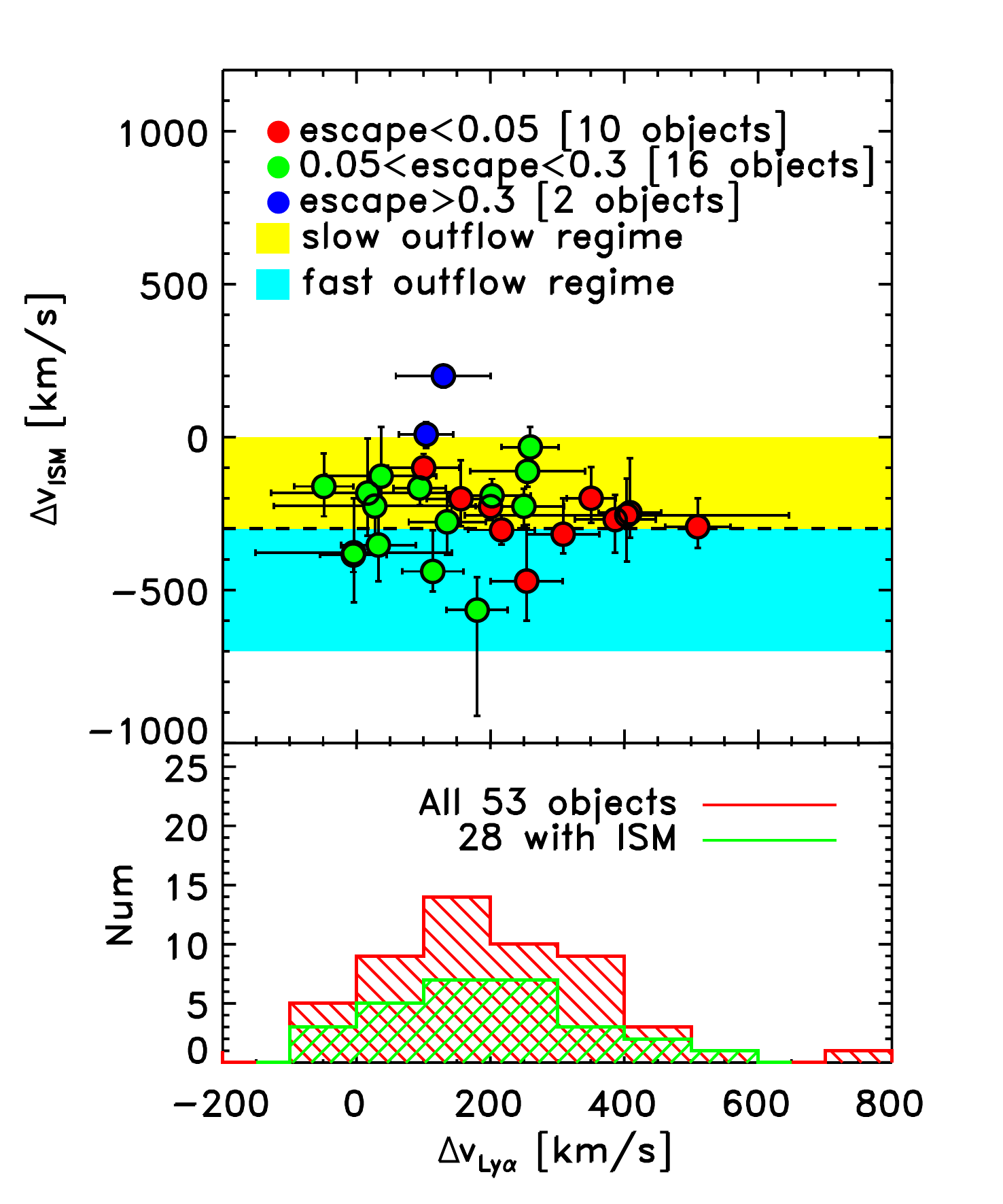}
   \caption{{\it Top panel:} Velocity difference between ISM lines and [CII] vs. velocity difference between \lya and [CII] for the 29 objects for which we were able to reliably estimate $z_{ISM}$. Red, green, and blue circles indicate objects with $f_{esc}(Ly\alpha)<0.05$, $0.05< f_{esc}(Ly\alpha)< 0.3$, and $f_{esc}(Ly\alpha)>0.3$, respectively. The yellow and cyan regions indicate the regimes of slow and fast outflows in the uniform expanding shell model by Verhamme et al. (2006) and Verhamme et al. (2015): for $v_{out}>$300 km s$^{-1}$ , the \lya offset is expected to always be small, regardless of the value of the outflow velocity; for $v_{out}<$300 km s$^{-1}$ , the \lya offset is determined by the neutral hydrogen column density NHI. {\it Bottom panel:} Histogram of the $\Delta v_{Ly\alpha}$ for all 53 galaxies in the sample (red) and for the 29 galaxies (green) for which we were able to determine $z_{ISM}$, which are shown in the upper panel.}
      \label{Fig:delta_vs_delta}
   \end{figure}

   \begin{figure*}[!ht]
   \centering
   \includegraphics[width=\textwidth]{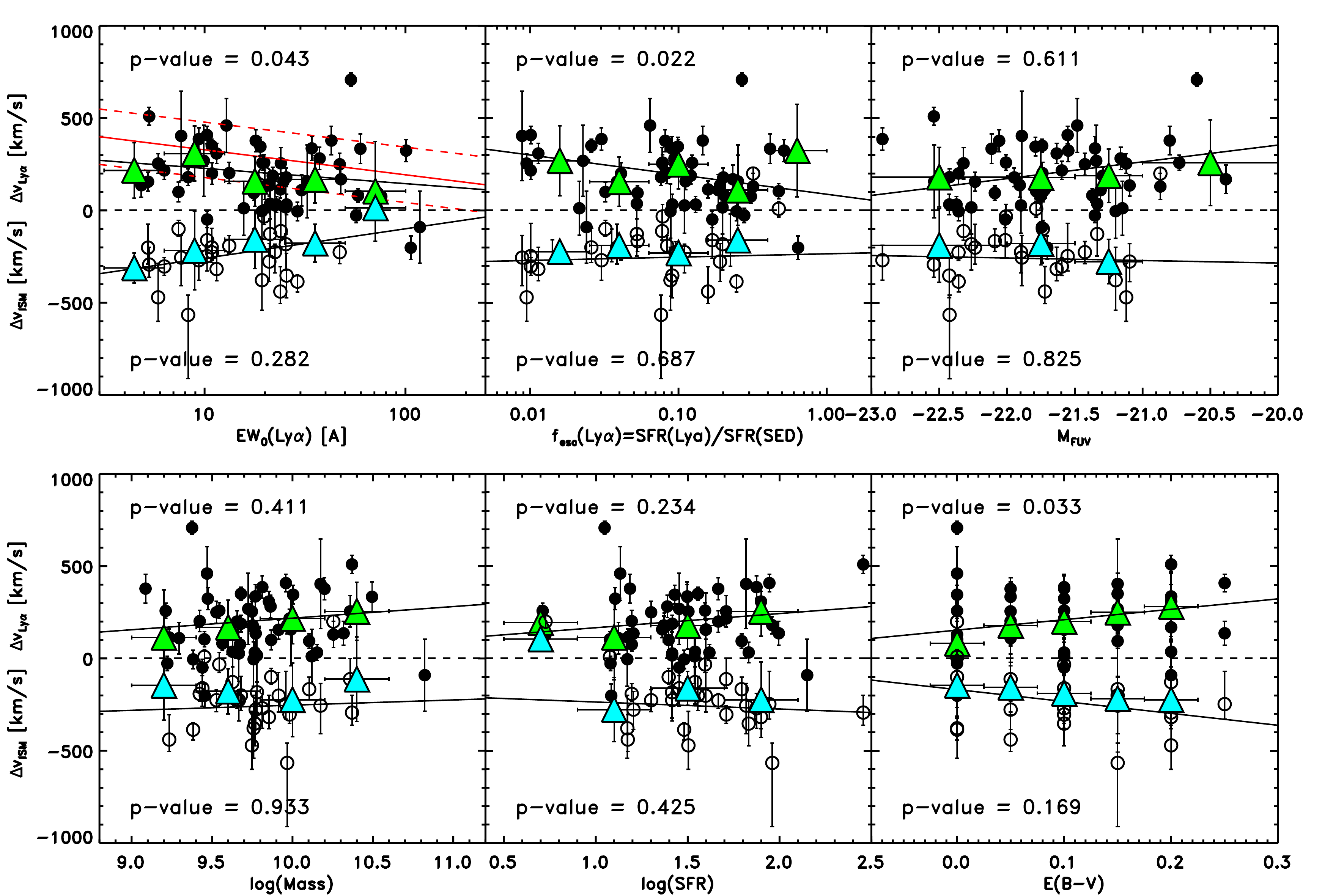}
   \caption{Velocity offset between \lya and [CII] (filled circles) and between ISM and [CII] (empty circles) as a function of $EW_0(Ly\alpha)$ (top left panel), $f_{esc}(Ly\alpha$) (top central panel), absolute magnitude in the far-UV M$_{FUV}$ (top right panel), stellar mass (bottom left panel), SFR (bottom central panel), and E(B-V)$_{\ast}$ (bottom right panel). The green (cyan) triangles show the running median of the \lya (ISM) offset in bins of $EW_0(Ly\alpha)$, $f_{esc}(Ly\alpha)$, redshift, stellar mass, SFR, and E(B-V)$_{\ast}$. The error bars on the offset velocity mainly reflect the spectral resolution of the optical spectra: smaller error bars are for objects from DEIMOS, which have $R\sim2000$, while larger bars are in general from VIMOS/VUDS spectroscopy, which has a lower spectral resolution (R$\sim$230). The continuous lines are linear fits to the black points, and the p-value of the Spearman rank test (the probability of no correlation) is reported in each panel.
   In the first panel ($\Delta v_{Ly\alpha}$ vs $EW_0(Ly\alpha$) we show in red the correlation between the two quantities reported in Erb et al. (2014): the continuous line is our best fit to the data in their Figure 6, and the dashed lines contain 68\% of their data points.
   }

      \label{Fig:correlation}
   \end{figure*}

Figure \ref{Fig:delta_vs_delta} shows the velocity offset $\Delta v_{ISM}$ as a function of the offset $\Delta v_{Ly\alpha}$  for the 29 galaxies for which a robust $z_{ISM}$ could be measured (i.e., objects with at least three detected ISM lines). No correlation between $\Delta v_{ISM}$ and $\Delta v_{Ly\alpha}$ is evident. When we interpret the blueshifted ISM lines as evidence for outflows, this implies that these outflows do not seem to strongly affect the \lya offset; in other words, larger or smaller \lya offsets can be found in galaxies with or without outflows. This is different from the results by Marchi et al. (2019), who found a correlation between these two quantities, with objects with small \lya offsets also having the largest (negative) ISM offsets (and vice versa). Interestingly, galaxies with different $f_{esc}(Ly\alpha)$ occupy quite different regions of the $\Delta v_{ISM}$ versus $\Delta v_{Ly\alpha}$ plane: regardless of their $\Delta v_{ISM}$, objects with low \lya escape fractions, $f_{esc}(Ly\alpha)<0.05$, tend to have larger \lya offsets ($\Delta v_{Ly\alpha}\gtrsim200$ km s$^{-1}$), while objects with intermediate \lya escape fractions, $0.05<f_{esc}(Ly\alpha)<0.3$, have smaller \lya offsets ($0\lesssim v_{Ly\alpha}\lesssim200$ km s$^{-1}$); the two objects with large \lya escape fractions, $f_{esc}(Ly\alpha)>0.3$ are also the only ones with positive $\Delta_{ISM}$ in this figure. Their redshifted ISM lines would indicate, formally, an inflow of gas in these galaxies. The first of them is a quite faint [CII] emitter, detected at only 4 $\sigma$, its [CII] spectrum is quite noisy, and [CII] may even show two velocity components; therefore it is possible in this case that the [CII] in this galaxy is not a reliable tracer of the systemic velocity. The other object with a high escape fraction, on the other hand, is very bright in [CII] (detected at 32 $\sigma$, the brightest emitter in the sample), and its [CII] emission is spatially extended: this galaxy might be embedded in a reservoir of cold gas that could be flowing in to fuel star formation.

The bottom panel of Figure~\ref{Fig:delta_vs_delta} also shows that galaxies with or without available ISM redshift determination have very similar $\Delta v_{Ly\alpha}$ distributions. In other words, the subsample of 29 galaxies with $z_{ISM}$ measurements is not different from the parent sample.

\subsection{Correlation of \lya offsets with other physical properties}
In Figure~\ref{Fig:correlation} we show the velocity offsets $\Delta v_{Ly\alpha}$ (for all 53 objects in the sample) and $\Delta v_{ISM}$ (for the subsample of 29 for which the ISM redshift could also be measured) as a function of $EW_0(Ly\alpha)$, $f_{esc}(Ly\alpha)$, absolute magnitude in the far-UV M$_{FUV}$, stellar mass, SFR, and E(B-V)$_{\ast}$. We ran a Spearman rank test to check and validate possible correlations between each parameter and the velocity offsets: in each panel we report the p-value, that is, the probability of the null hypothesis (no correlation) being true, ranging from 0 to 1: a high p-value means that there is no correlation, and a low p-value confirms the correlation.

We find that the \lya velocity offsets  are anticorrelated with $EW_0(Ly\alpha)$ and $f_{esc}(Ly\alpha$), but they correlate positively with the E(B-V)$_{\ast}$: in these cases, the p-values are lower than 0.05. In particular, galaxies with smaller $EW_0(Ly\alpha)$ (or equivalently, smaller $f_{esc}(Ly\alpha)$) on average have higher $\Delta v_{Ly\alpha}$ than galaxies with larger $EW_0(Ly\alpha)$ (or $f_{esc}(Ly\alpha$)): this agrees perfectly well with the results shown in Figure~\ref{Fig:delta_vs_delta}, although this figure only shows the 29 out of 53 objects for which $\Delta v_{ISM}$ could also be measured. Erb et al. (2014) found a similar correlation between $\Delta v_{Ly\alpha}$ and  $EW_0(Ly\alpha)$ at $z\sim 2-3$ (highlighted in red in Figure~\ref{Fig:correlation}); Du et al. (2018) confirmed that this correlation does not evolve and still holds up to $z\sim4$ (although they used composite spectra, while we worked with individual measurements); Shapley et al. (2003), although using $\Delta(Ly\alpha-ISM)$, find similar correlations between offset velocity and $EW_0(Ly\alpha)$ (or E(B-V)$_{\ast}$). It is interesting to note that the relation we find at $z\sim5$ between $\Delta v_{Ly\alpha}$ and  $EW_0(Ly\alpha)$ is not exactly the same Erb et al. (2014) found at $z\sim2-3$; however, this relation  does not seem to evolve very significantly from $z\sim5$ and $z\sim2-3$: the intercepts of the two linear fits agree within 1$\sigma$, and the slopes are different only at a 2$\sigma$ level.

In addition, we find that objects with more reddening have on average larger $\Delta v_{Ly\alpha}$: this is expected considering the correlation between $\Delta v_{Ly\alpha}$ and  $EW_0(Ly\alpha)$ discussed above at the known correlation between E(B-V)$_{\ast}$ and $EW_0(Ly\alpha)$ (Pentericci et al. 2007, Shapley et al. 2003, Kornei et al. 2010, Hayes et al. 2014, Cassata et al. 2015). No significant correlations are observed between $\Delta v_{Ly\alpha}$ and stellar mass, SFR, and M$_{FUV}$: the Spearman rank test gives probabilities for the null hypothesis to be true above 40\%. This agrees with findings by Steidel et al. (2010), who found no significant correlation between $\Delta v_{Ly\alpha}$ and stellar mass either for their sample of star-forming galaxies at $z\sim2-3$.

Although we do not report this in Figure~\ref{Fig:correlation}, we also find no correlation  between offset velocities (\lya or ISM) with redshift within our sample: the probability that no correlation is true is $\sim$50\% for the ISM offsets and $\sim$80\% for the \lya according to the Spearman test. As mentioned in Section~2, the redshift coverage of the ALPINE sample is not continuous: about two-thirds of the objects are at $4.4<z<4.6$ and one-third is at $5<z<6$; both redshift bins have a similar median $\Delta v(Ly\alpha) \sim200$ km s$^{-1}$. At the same time, we do not see any robust correlation between $EW_0(Ly\alpha)$, \lya escape fraction, M$_{FUV}$, stellar mass, SFR, and E(B-V)$_{\ast}$ with $\Delta v_{ISM}$: the probability that no correlation is present between these quantities and $\Delta v_{ISM}$ is above 20\% in all cases.

\section{Discussion}

While it is not straightforward to interpret $\Delta v_{Ly\alpha}$ in terms of internal physical properties of galaxies (the observed offset is a complicated function of the presence and velocity of outflowing gas, geometry, and the distribution of gas and neutral gas column density NHI, see, e.g., Verhamme et al. 2015), the physical process producing the absorption of ISM lines is quite simple, and the prevalence of blueshifted ISM lines in star-forming galaxies (and in our sample in particular) can be interpreted as evidence for outflows. Steidel et al. (2010), Du et al. (2018), and Steidel et al. (2018), among others, found that ISM lines are often broad and saturated, spanning velocity intervals between 0 and -1000 km s$^{-1}$. This indicates that gas is outflowing with a range of velocities, from gas that is at the systemic velocity of the galaxy up to gas that escapes at 800-1000 km s$^{-1}$. Unfortunately, the S/N of our spectra is not enough to analyze the details of the ISM line morphologies, and we can only estimate the centroid of the line. In the following, we assume for simplicity that the $\Delta v_{ISM}$, estimated from the centroid of the ISM lines, traces the $v_{out}$ (although we have to keep in mind that the gas is very probably outflowing with a range of velocities, with faster and slower components, and therefore the relation between $\Delta v_{ISM}$ and $v_{out}$ might be in reality less simple, see Steidel~et~al.~2010 and Martin et al. 2012). The centroids are almost always blueshifted, which implies the presence of outflowing gas in the bulk of the galaxies in our sample, with average velocities $v_{out}\sim200-500$ km s$^{-1}$ (but we have to keep in mind that surely there are even faster components). This is consistent with the results for the whole ALPINE sample by Ginolfi et al. (2020), who analyzed the spectral shape of the [CII] emission to identify outflows in the most highly star-forming galaxies in the sample.

The interpretation of \lya emission is far more complicated. The uniform expanding shell model (e.g., Verhamme et al. 2006, Verhamme et al. 2008, Verhamme et al. 2015, Gronke, Bull \& Dijkstra 2015) is widely used and quite successful in reproducing the spectral profiles of \lya for star-forming galaxies at all redshifts (e.g., Hashimoto et al. 2015, Duval et al. 2016, Gronke 2017, Orlitova et al. 2018, Marchi et al. 2019). In its framework, the region that emits \lya photons is surrounded by a uniform shell of gas that expands at a fixed velocity, and the variety of spectral shapes for the \lya line in star-forming galaxies can be reproduced by varying only six parameters (Gronke 2017). This model might be simplistic for the galaxies in our sample, which have continuous and smooth star formation (Faisst et al. 2020), often display small but significant spatial offsets between the \lya and the UV continuum emission (indicating that the ISM is likely not homogeneous), and have mixed morphologies (see Appendix~\ref{appendix}).   Gronke~\&~Dijkstra~(2016) suggested that extracting information from shell models is not straightforward, and they cautioned against overinterpreting the results from these models; nevertheless, Gronke et al. (2016) offered a possible explanation for the success of the uniform shell model in reproducing the \lya line spectral morphology, showing that the solution for a very clumpy medium essentially converges to the uniform case. 

According to the  uniform shell model, galaxies with fast outflows have in general smaller \lya offsets: when the \lya photons reach the fast expanding shell, they are seen out of resonance and not absorbed; on the other hand, \lya photons in galaxies with slower outflows do not take advantage of this effect, and are absorbed more or less easily at resonance (at the systemic velocity of the system) depending on the neutral hydrogen column density NHI. In this respect, it is worth noting that in our sample we have 20 of 29 galaxies in the {\it \textup{slow}} outflow regime (v$_{out}<$ 300 km s$^{-1}$), for which the shift of the \lya line depends strongly on NHI, and 8 of 29 in the {\it \textup{fast}} outflow regime (v$_{out}>$ 300 km s$^{-1}$): most of these latter (5 out of 8) have a large \lya escape fraction and small $\Delta v_{Ly\alpha}$, which qualitatively agrees with the predictions of the uniform expanding model. It is worth noting that of all the sources with slow outflows and $\Delta v_{Ly\alpha}>300$ km s$^{-1}$, 4 have very low escape fractions. This is consistent with the fact that according to this model, very large shifts of the \lya line in galaxies with slow outflows can be observed only for very large NHI, log(NHI[cm$^{-2}$])$>$ 20.  Galaxies with $\Delta v_{Ly\alpha} <300$ km s$^{-1}$ are a mixture of fast and slow outflows, and most of them have intermediate escape fractions. Sources with $100<\Delta v_{Ly\alpha} <300$ km s$^{-1}$ would in the framework of the expanding shell model have an average log(NHI[cm$^{-2}$])$ \sim$ 19.5, which is lower than for sources with $\Delta v_{Ly\alpha}>300$ km s$^{-1}$, consistently with their characteristic $f_{esc}(Ly\alpha)$ values. Finally, galaxies with $\Delta v_{Ly\alpha} <100$  km s$^{-1}$ {would} span a wide range of NHI values, from log(NHI[cm$^{-2}$]) = 16 to log(NHI[cm$^{-2}$])= 21, therefore we cannot make tentative conclusions on their average log(NHI[cm$^{-2}$]) value. It is important to keep in mind that these average NHI values are obtained in the framework of a model that assumes a uniform composition for the expanding material. Allowing for a decreasing covering fraction, for the same total NHI, would also lead to smaller \lya velocity offsets and larger $EW_0(Ly\alpha)$ (e.g., Reddy et al. 2016; Steidel et al. 2018). 

Steidel et al. (2010) presented an alternative phenomenological model to interpret the spectroscopic data in their sample. This was physically motivated by the observation that the blueshifted ISM lines in the spectrum of $z\sim2-3$ galaxies are also broad and saturated, with a range of covering fractions of the underlying stellar continuum source. According to this model, the ISM gas absorbing the stellar continuum is outflowing at a range of velocities, with the outflow velocity a monotonic function of the galactocentric distance. This model can simultaneously reproduce the broad ISM blueshifted absorption lines and the asymmetric redshifted \lya line in Steidel et al. (2010). The offset of the \lya line in this case is the result of the combination of the absorption due to the gas near the systemic redshift of the galaxy and the velocity of the outflowing gas that ultimately scatter the \lya photons towards us. This model can reproduce the complex spectral morphologies of the \lya line: when there is less (more) NHI near the systemic velocity in the galaxy, the model predicts small (large) \lya offsets and strong (weak) Ly$\alpha$, naturally predicting the anticorrelation between $\Delta v_{Ly\alpha}$ and $EW_0(Ly\alpha)$ observed by many others and ourselves. Although the S/N of our spectra are not good enough to allow for a detailed study of the shape of the ISM absorption lines in comparison with Ly$\alpha$, which would be needed to fully test this model, we can only confirm zeroth-order predictions of this prescription. In particular, this model predicts a quite consistent $\Delta v_{Ly\alpha}$ - $\Delta v_{ISM}$ difference, in the sense that when $\Delta v_{ISM}$ moves toward more positive velocities because more NHI is present near the systemic velocity, $\Delta v_{Ly\alpha}$ also shifts redward. Our sample does not seem to show this behavior, as we do not see any positive correlation between $\Delta v_{ISM}$ and $\Delta v_{Ly\alpha}$ (see Fig.~\ref{Fig:delta_vs_delta}), although it is possible that our large error bars wash out the signal.

   \begin{figure}[!h]
   \centering
   \includegraphics[width=\columnwidth]{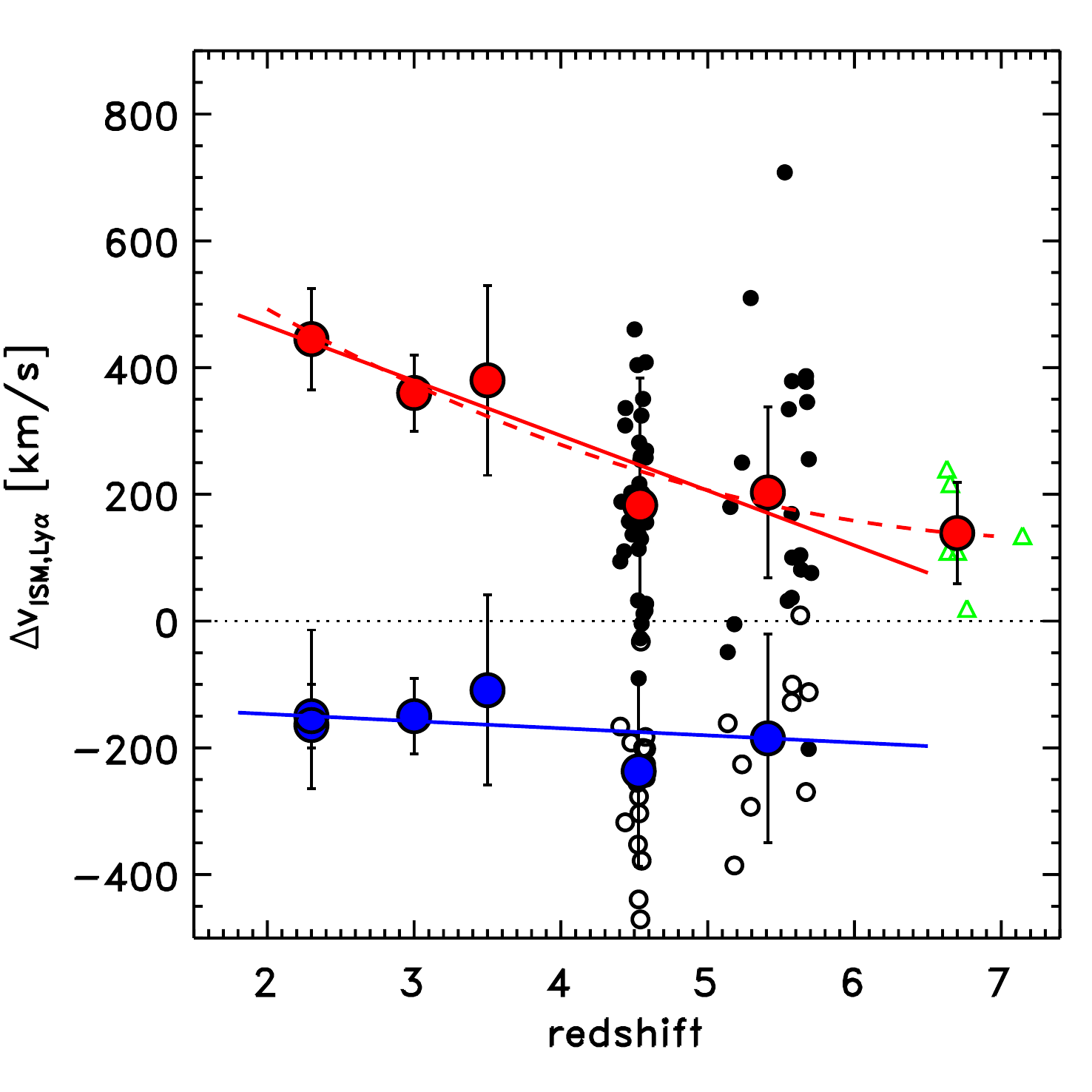}
   \caption{Evolution of the \lya and ISM offsets with redshift. We report individual measurements for our sample (filled and empty symbols indicate $\Delta_{Ly\alpha}$ and $\Delta_{ISM}$, respectively), and for galaxies at $z\sim7$ (green triangles, from Pentericci et al. 2016; Brada\v{c} et al. 2017; Matthee et al. 2019). Red and blue circles indicate the average values of $\Delta_{Ly\alpha}$ and $\Delta_{ISM}$ for different samples at different redshifts: the values at $z\sim2.3$ are from Steidel et al. (2010); at $z\sim2.3$ (for the ISM alone) they are from Talia et al. (2017); at $z\sim3$ they are from the stacked spectrum in Shapley et al. (2003); at $z\sim3.5$ they are from Marchi et al. (2019); at $z\sim4.5$ and 5.5 they are from this work; and at $z\sim7$ they are for \lya alone from Pentericci et al. (2016), Brada\v{c} et al. (2017), and Matthee et al. (2019). The blue and red continuous lines are linear best fits to the points. In the bottom panel we also report second-degree polynomial fit that also incorporates the $z\sim7$ data well.}
      \label{Fig:delta_z}
   \end{figure}

\section{Conclusions}
We used the [CII] line, detected in 53 star-forming galaxies at $4.4<z<6$ that were observed as part of ALPINE, to trace the systemic velocity of these galaxies and to measure the velocity offsets of \lya (and ISM lines, for a subsample of 29 galaxies) with respect to it. We then used these offsets to gain insights into the ISM properties of galaxies and to understand the physical driver of the increase in the escape fraction of \lya photons that is observed from $z\sim2$ to $z\sim6$ (Stark et al. 2010; 2011; Schenker et al. 2012; Cassata et al. 2015).

We stress that the original selection of the parent ALPINE sample (see Le F\`evre et al. 2020, B\'ethermin et al. 2020, Faisst et al. 2020) is primarily based on the UV-optical rest-frame properties of these galaxies: the targets are selected to be main-sequence galaxies at $4.4<z<6$, and therefore they can be considered typical galaxies at such epochs. They indeed have stellar masses similar to samples in the literature at lower redshifts, such as those in Steidel et al. (2010), Erb et al. (2004), Du et al. (2018), Pahl et al. (2020), and Marchi et al. (2019), with higher SFRs and brighter $M_{FUV}$ according to the relative evolution of the star-forming main sequence from lower redshift to $z\sim5$. We also showed in Section~\ref{Section:sample_data} that the requirement of accurate and robust spectroscopic redshifts, which is required to appropriately tune the ALMA spectral setting to cover the [CII] line, did not bias our sample toward strong \lya emitters. As a result, this is the largest sample of main-sequence star-forming galaxies at $z>4$ with published $\Delta v_{Ly\alpha}$ and $\Delta v_{ISM}$ offsets, and one of the largest samples at all redshifts.

The main result of the paper is that the bulk of our galaxies show small $\Delta v_{Ly\alpha}$ offsets, compared with other works at $z\sim2-3.5$, with a peak around 200 km s$^{-1}$ and up to $\sim$ 400 km s$^{-1}$; at the same time, we also measure $\Delta v_{ISM}$ offsets between -500 and 0 km s$^{-1}$ (see Figure~\ref{Fig:hist_delta}). We do not find any significant correlation between $\Delta v_{Ly\alpha}$ and $\Delta v_{ISM}$, but we find a correlation between $\Delta v_{Ly\alpha}$ and  $f_{esc}(Ly\alpha)$ (or, equivalently, $EW_0(Ly\alpha)$, as found at lower redshift among others by Erb et al.~2014 and Du et al. 2018). These correlations are highlighted in the first two panels of Figure~\ref{Fig:correlation}, and are also evident in  Figure~\ref{Fig:delta_vs_delta}: objects with smaller $f_{esc}(Ly\alpha)$ tend to have larger  $\Delta v_{Ly\alpha}$, and vice versa. 

We also explored the possible correlation between the offset velocities Ly$\alpha$-[CII] and ISM-[CII] and a number of properties, such as $EW_0(Ly\alpha)$, \lya escape fraction, stellar mass, SFR, E(B-V)$_{\ast}$, and $M_{FUV}$. The strongest and more robust correlations we found are between \lya escape fraction, $EW_0(Ly\alpha)$, and E(B-V)$_{\ast}$ with the \lya offsets: these correlations are routinely observed at lower redshift (Erb et al. 2014; Trainor et al. 2015; Pentericci et al. 2007, Shapley et  al. 2003, Kornei et al. 2010, Hayes et al. 2014) and can be easily explained by models (uniform expanding shell or more complex configurations such as in Steidel et al. 2010): objects with large $f_{esc}(Ly\alpha$) (and consequently larger $EW_0(Ly\alpha)$) are those with smaller \lya offsets because they are those with less dust (smaller E(B-V)$_{\ast}$), smaller NHI, or smaller gas covering fraction. 
We do not observe a large evolution of the correlation between $\Delta v_{Ly\alpha}$ and $EW_0{Ly\alpha}$ between $z\sim2-3$ (Erb et al. 2014), and $z\sim5$ (this paper; see Figure 4 and Section 3): we only find a 2$\sigma$ difference for the slope of the relation, while the two intercepts agree to within 1$\sigma$.

At the same time, no strong correlations between physical properties and ISM offsets are found. When we assume that the ISM offsets are the direct evidence of outflows, these results imply that at least at these redshifts, outflows are not the main driver of the escape of \lya photons. This is because, returning to the analysis of Figure~\ref{Fig:delta_vs_delta}, the outflow velocities in most of the galaxies in our sample are quite slow (in the slow-outflow regime for the uniform expanding shell model) and the \lya photons are therefore easily absorbed by the gas at the systemic velocity of the galaxies. Coincidentally, Figure~\ref{Fig:correlation} also clearly shows that objects  with fast outflow velocities ($v_{out}>300$ km s$^{-1}$, or $\Delta v_{ISM}< -300$ km s$^{-1}$) tend to have large $f_{esc}(Ly\alpha$) (top middle panel), as expected from all models, but they are not numerous enough to drive a correlation between $\Delta v_{ISM}$ and \lya escape fraction.

Finally, it is interesting to compare the velocity offsets $\Delta v_{Ly\alpha}$ and $\Delta v_{ISM}$ for our sample at $z\sim5$  with other similar samples in the literature at lower redshifts. In Figure~\ref{Fig:delta_z} we report the average and dispersion of these offsets for our sample in comparison with the samples by  Steidel~et~al.~(2010) at $z\sim2.3$, Talia et al. (2017) at $z\sim2.3$, Shapley et al. (2003) at $z\sim3$ (a stack of 800 spectra), and Marchi et al. (2019) at $z\sim3.5$, and with the handful of measurements for \lya only at $z\sim7$ by Pentericci et al. (2016), Brada\v{c} et al. (2017) and Matthee et al. (2019). The ISM offsets do not vary significantly from $z\sim2$ to $z\sim5$: the linear best fit has a negative slope, but the slope is compatible with zero within the errors. At the same time, $\Delta v_{Ly\alpha}$ significantly evolves over the same redshift interval, it extends from $\sim500$ km s$^{-1}$ at $z\sim2$ to $\sim200$ km s$^{-1}$ $z\geq5$, with a negative slope that is different from zero at 4$\sigma$. When we consider the observed evolution of the fraction of LAE emitters with redshift (Stark et al. 2011; Mallery et al. 2012; Cassata et al. 2015) together with the well-established correlation between $EW_0(Ly\alpha$) and $\Delta v_{Ly\alpha}$ (first panel of Figure~\ref{tab:properties_photo1}, Shapley et al. 2003, Erb et al. 2014), it is indeed expected to find that the average $\Delta v_{Ly\alpha}$ decreases with redshift: at increasing redshift, between $z\sim2$ and $z\sim6$, samples of normal star-forming galaxies contain increasingly more strong \lya emitters (meaning that the distribution of $EW_0(Ly\alpha$) shifts toward higher values), and therefore, through the correlation between $\Delta v_{Ly\alpha}$ and $EW_0({Ly\alpha}$), these samples display smaller \lya velocity offsets on average. 

These results together can give interesting insights that help to interpret the observed increase in the fraction of \lya emitters in the same redshift interval (Schenker et al. 2012, Cassata et al. 2015), or similarly, the observed increase in the average \lya escape fraction (Hayes et al. 2011; Konno et al. 2016). If the average ISM offsets do not evolve, this would imply that the incidence of outflows, and their velocity, does not change significantly from $z\sim2$ to $z\sim5$; assuming therefore no significant change in the incidence of fast and slow outflows over that redshift range (again: the peak and dispersion do not change significantly), the observed decrease of $\Delta v_{Ly\alpha}$ in the framework of the uniform expanding shell model might be due to a decrease in the amount of NHI (and possibly dust reddening) from $z\sim2$ to $z\sim7$. Konno et al. (2016) suggested that \emph{\textup{in situ,}} NHI decreases by almost one order of magnitude from $z\sim2$ to $z\sim6$ as a result of a simultaneous increase in the ionization parameter, which is consistent with our interpretation. Alternatively, models that do not assume a uniform composition for the expanding material (e.g., Steidel et al. 2010; 2018) show that the observed increasing fraction of strong LAE with redshift and the simultaneous decrease in $\Delta v_{Ly\alpha}$ might also be explained by a decrease in gas covering fraction, or in other words, by a more clumpy distribution of gas and dust (Du et al. 2018; Pahl et al. 2020). If this trend of decreasing NHI, dust content, and/or gas-covering fraction continues beyond $z\sim6$ into the epoch of reionization, this would also imply an increase in the escape fraction of Lyman continuum photons that could make up the bulk of the ionizing continuum.

\begin{acknowledgements}
This paper is dedicated to the memory of Olivier Le F\`evre, PI of the ALPINE survey. This paper is based on data obtained with the ALMA observatory, under the Large Program 2017.1.00428.L. ALMA is a partnership of ESO (representing its member states), NSF (USA) and NINS (Japan), together with NRC (Canada), MOST and ASIAA (Taiwan), and KASI (Republic of Korea), in cooperation with the Republic of Chile. The Joint ALMA Observatory is operated by ESO, AUI/NRAO and NAOJ. Based on data products from observations made with ESO Telescopes at the La Silla Paranal Observatory under ESO programme ID 179.A-2005 and on data products produced by CALET and the Cambridge Astronomy Survey Unit on behalf of the UltraVISTA consortium. A significant fraction of the spectrographic data presented herein were obtained at the W.M. Keck Observatory, which is operated as a scientific partnership among the California Institute of Technology, the University of California, and the National Aeronautics and Space Administration. The Observatory was made possible by the generous financial support of the W.M. Keck Foundation. We thank the indigenous Hawaiian community for allowing us to be guests on their sacred mountain, a privilege, without which, this work would not have been possible. We are most fortunate to be able to conduct observations from this site.

PC and LM acknowledge support from the BIRD 2018 research grant from the Universit\`a degli Studi di Padova. PC, AC, CG, FL, FP, MT, GR acknowledge the grant PRIN MIUR 2017. JDS was supported by JSPS KAKENHI Grant Number JP18H04346, and the World Premier International Research Center Initiative (WPI Initiative), MEXT, Japan. DAR acknowledges support from the National Science Foundation under grant numbers AST-1614213 and AST-1910107 and from the Alexander von Humboldt Foundation through a Humboldt Research Fellowship for Experienced Researchers. EI acknowledges partial support from FONDECYT through grant N$^\circ$\,1171710

The authors thank Anne Verhamme for useful discussions that helped improving this manuscript, and the anonymous referee for the very constructive and insightful report.
\end{acknowledgements}

\begin{appendices}\label{appendix}
\section{Coherence of regions emitting Ly$\alpha$ and [CII]}
\begin{figure*}
   \centering
   \includegraphics[trim={0 1.5cm 0 1cm}, clip, width=0.96\textwidth ]{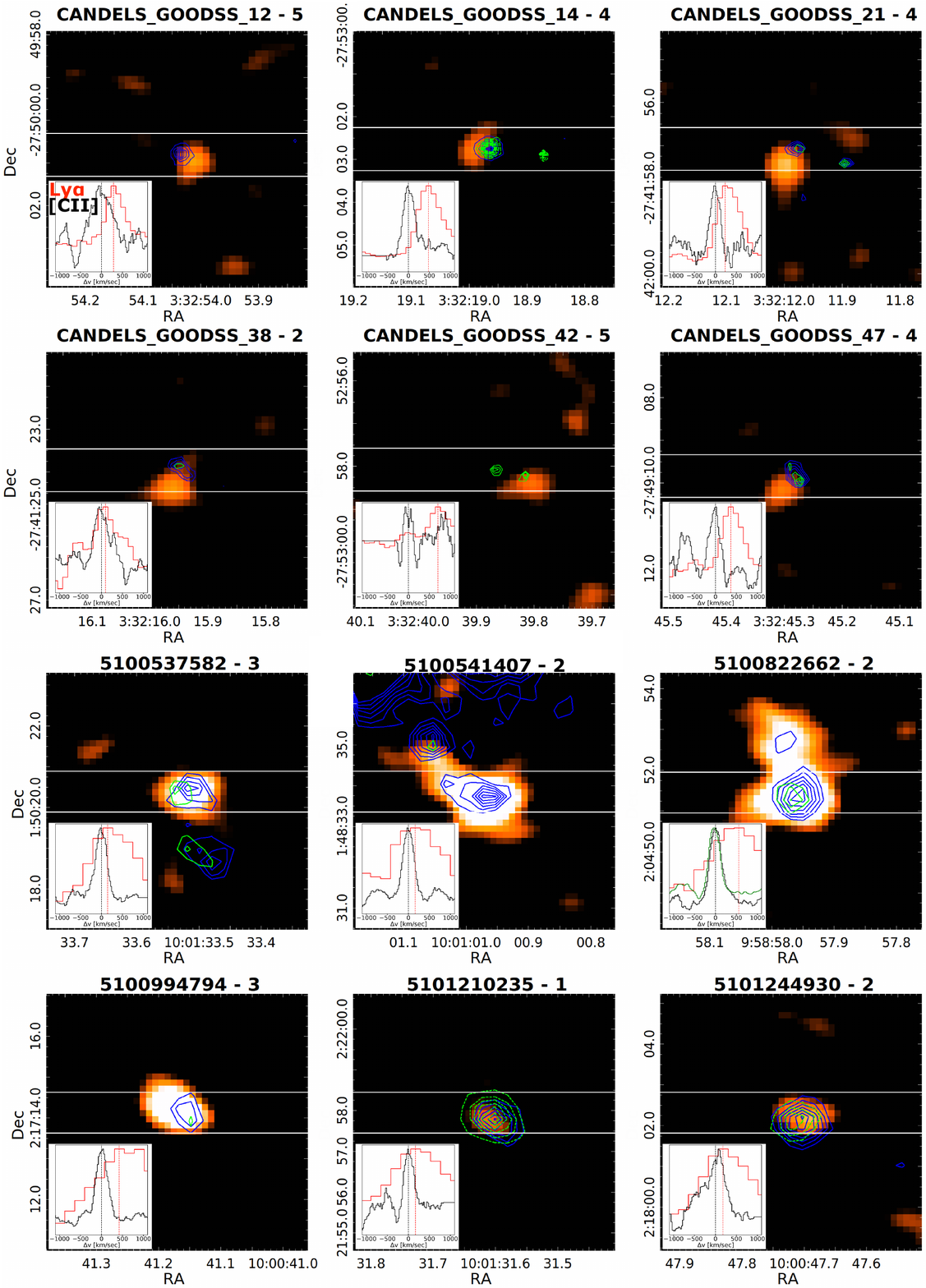}
   \caption{Velocity-integrated [CII] flux maps. The green contours represent the emission in the filter where the \lya line falls in: {\it i+} or {\it r+} SUBARU bands, depending on the redshift of the source. The blue contours mark the emission in the ULTRAVISTA Ks filter. Solid contours are drawn at steps of 1$\sigma$, while dashed contours are at steps of 2$\sigma$, both starting from 3$\sigma$. The white lines mark the position of the spectral slit. The plot in the bottom left corner represents the [CII] (black) and \lya (red) lines. The code next to the object name in the top label indicates the morpho-kinematic class from Le F\`evre et al. (2020): 1.0 are rotators; 2.0 are mergers; 3.0 are extended [CII] emitters, dispersion dominated; 4.0 are compact, dispersion dominated; and 5.0 are faint [CII] emitters.}
      \label{Fig:cutouts1}
   \end{figure*}

In this appendix we compare the spatial distributions of the regions emitting [CII] with those emitting Ly$\alpha$ for our 53 objects. Figure~\ref{Fig:cutouts1} shows the [CII] map, overlaid with the contours of the optical image containing the Ly$\alpha$ line (depending on the redshift, $i_+$ or $r_+$ SUBARU images for the COSMOS field, and ACS/F814W image for the GOODS-S field) and an image that is as similar as possible to the optical rest-frame (UltraVISTA DR4 $K_s$ image for the COSMOS field, and WFC3/F160W for the GOODS-S field), and the position of the spectroscopic slit, which is always 1" wide. We confirmed that for each optical spectrum, the Ly$\alpha$ emission is always aligned within 1" (typically, the spatial resolution of the spectroscopic observations) with the UV rest-frame continuum, and therefore it can be considered spatially coincident, within the same precision, with the position of the optical photometric emission. We also show the spectra around [CII] and Ly$\alpha$ lines, centered in velocity on the position of [CII]. The [CII] spectra shown here are obtained by integrating over the regions in which the [CII] emission is detected at $>2\sigma$ (see B\'ethermin et al. 2020 for details); for cases in which the [CII] emission is more extended than the optical, or for which the [CII] emission has many components, we re-extracted a [CII] spectrum from a smaller region coincident with the \lya/optical region to determine whether the [CII] velocity peak moves. We also report this spectrum in the insets. 

We also report in the label of each panel the morpho-kinematic class estimated in Le F\`evre et al. (2020) according to the following coding: 1. rotating disks; 2. mergers; 3. extended, dispersion dominated; 4. compact, dispersion dominated; and 5. too faint to be classified. Six of the 53 galaxies studied here are disks, 20 are mergers, 14 are dispersion dominated (extended or compact), and 8 are faint. The most frequently represented class (40\%) are mergers: objects are classified as such in three cases: i) the [CII] image shows multiple components; ii)  multiple components appear in the position-velocity diagram extracted from the [CII] cube; and iii) the ancillary images, typically for the COSMOS field HST/ACS (Koekemoer et al. 2007) and for the GOODS-S field HST/WFC3 (Koekemoer et al. 2011), show multiple components.
We comment on the result of this analysis below for each object.
\begin{itemize} 
     \item {\bf Object 12}: this object is too faint in [CII] for us to perform a morpho-kinematic classification. The \lya-optical light is offset by $\sim$0.5" from the peak of the [CII] emission. There is no evidence of multiple components.
    
    \item {\bf Object 14}: this object is classified as dispersion dominated. The \lya-optical light is offset by $\sim$0.5" from the peak of the [CII] emission.
    
    \item {\bf Object 21}: this object is classified as dispersion dominated. The \lya-optical light is offset by $\sim$0.5" from the peak of the [CII] emission.
    
    \item {\bf Object 38}: this object is classified as a merger because multiple components, separated by $\sim$250 km s$^{-1}$, are present in the position-velocity diagram (see Le F\`evre et al. 2020). The \lya-optical light is offset by $\sim$0.5" from the peak of the [CII] emission.
    
    \item {\bf Object 42}: this object is too faint in [CII] to perform a morpho-kinematic classification. The optical image close to \lya (in green in the figure) reveals two weak peaks, and the \lya line is extracted from the right peak, closest to the [CII] peak.
    
    \item {\bf Object 47}: this object is classified as dispersion dominated. The \lya-optical rest-frame emission is offset by $\sim$0.5" from the peak of the [CII] map. 

    \item {\bf Object 5100537582}: this object is classified as dispersion dominated. The\lya-optical emission is coincident with the peak of the [CII] map. There is no evidence of multiple components.
    
    \item {\bf Object 5100541407}: this object is classified as a merger because the [CII]  map shows multiple bright clumps. The UltraVISTA $K_s$ image (mapping the optical rest-frame light) also shows two components, but only one of them is in the spectroscopic slit. It is clear that the \lya spectrum has been extracted from the region coincident with the brightest [CII] peak, and we verified that the position of the [CII] velocity peak is not affected by the faintest clumps.
    
    \item {\bf Object 510082662}: this object is classified as a merger because the [CII] emission originates from two bright components. However, only one of these two [CII] clumps is coincident with the UV-optical emission, where the \lya spectrum has been extracted.

    \item {\bf Object 5100994794}: this object is classified as dispersion dominated. The [CII] emission is quite extended and well centered in the spectroscopic slit. The \lya and [CII] spectra are extracted from the same physical region.
 
    \item {\bf Object 5101210235}: this object is classified as a rotator. The \lya-optical emission is coincident with the peak of the [CII] emission.
    
    \item {\bf Object 5101244930}: this object is classified as a merger because multiple clumps appear in the ancillary HST and Subaru optical images. The optical rest-frame light, traced by the UltraVISTA $K_s$ image, is coincident with the peak of the [CII] emission, while the \lya-UV light might be slightly offset from it, by 0.5" at most. 
   
    \item {\bf Object 510605533}: this object is too faint in [CII] for us to perform a morpho-kinematic classification. The \lya-optical rest-frame light is offset by $\sim$0.5" from the peak of the [CII] emission.
    
    \item {\bf Object 510786441}: this object is classified as a merger because multiple components are detected in the position-velocity diagram (Le F\`evre et al. 2020), separated by $\sim$200 km s$^{-1}$. The \lya-optical emission is spatially coincident with the peak of the [CII] emission.
    
    \item {\bf Object 5180966608}:  this object is classified as a merger because multiple clumps appear in the ancillary HST and Subaru optical images, although the [CII] emission is extended and is not resolved in multiple components. The \lya line has been extracted from the UV-optical peak that is coincident with the [CII] emission. Because [CII] is extended, we re-extracted a [CII] spectrum from a region coincident with the UV-optical region, and we find that the position of the [CII] velocity peak is not affected.

   \item {\bf Object 274035}: the optical emission (tracing the \lya emission and UV rest-frame continuum) and the $K_s$ emission (tracing close to the optical rest-frame) are offset by $0.5$" and $0.3$" from the peak of the [CII] emission. The object is classified as dispersion dominated, and there are no signs of multiple components from the images or from the spectra.
    
    \item {\bf Object 308643}: this object is classified as a merger mainly because two dynamical components are present in the position-velocity diagram (see Le F\`evre et al. 2020), although a rotating disk could also produce a similar velocity pattern. The spatial UV-[CII] and spectral \lya-[CII] offsets are both very small.
    
    \item {\bf Object 351640}: this object is classified as a dispersion-dominated galaxy. It is very weak both in the optical and in the NIR image, which is expected because it is selected as \lya emitter by the narrow-band technique, and therefore it is difficult to identify possible spatial offsets between \lya and [CII]. There is no evidence of multiple components.
    
    \item {\bf Object 372292}: this object is classified as a merger mainly because two dynamical components are present in the position-velocity diagram, separated in velocity by $\sim$ 200 km s$^{-1}$ (see Le F\`evre et al. 2020), although a rotating disk could also produce a similar velocity pattern. Neither the spatial nor the spectral \lya-[CII] offset are significant.
    
    \item {\bf Object 400160}: this object is classified as dispersion dominated. There is no evidence of multiple components, and there is no significant offset between the UV and optical rest-frame emission and [CII].
    
    \item {\bf Object 403030}: this object is classified as a merger mainly because the optical SUBARU images, tracing the UV rest-frame, show multiple components (green contours in the figure). The \lya spectral profile is double peaked (the main peak lies at $\sim$ +400 km s$^{-1}$, the other at $\sim$ -200 km s$^{-1}$), but we confirmed in the 2D spectrum that there is no spatial offsets between the two peaks, meaning that within the resolution, they arise from the same region. Double-peaked \lya profiles are easily predicted in the framework of the homogeneous expanding shell model (Verhamme et al. 2008).
    
    \item {\bf Object 416105}: this object is classified as a rotator. Small and insignificant spatial offsets are present between the UV and optical rest-frame and the [CII] emission. There is no evidence of multiple components.
  
    \item {\bf Object 417567}: this object is classified as a merger because the [CII] maps show fainter components close to the main central source. The morphology in the UV and optical rest-frame is disturbed by the presence of a nearby foreground source, nevertheless, it is clear that the \lya emission originates from the main [CII] component. We confirmed that the secondary [CII] peaks do not contribute to moving the position of the [CII] velocity peak.
    
    \item {\bf Object 422677}: this object is classified as a merger because multiple kinematic components are present in the position velocity diagram, separated in velocity by less than 200 km s$^{-1}$ (see Le F\`evre et al. 2020). However, the [CII] emission is coincident with the optical rest-frame light and is only slightly offset from the \lya-UV rest-frame emission. 
    
    \item {\bf Object 430951}: this object is too faint in the [CII] map to be morpho-kinematically classified. The [CII] spectrum is quite wide, with a red tail extending up to +1500 km s$^{-1}$ from the spectral peak. Moreover, the [CII] emission is offset from the position of the spectroscopic slit, therefore [CII] and \lya  in this case do not originate from the same physical region. Interestingly, this is one of the few cases in which we measure a negative \lya-[CII] velocity offset.
    
    \item {\bf Object 432340}: this object is classified as a merger because multiple components are present in the [CII] map. It is clear, however, that the \lya emission (and the optical rest frame) originate from the main [CII] clump, and the secondary [CII] clumps do not affect the position of the [CII] velocity peak.
    
    \item {\bf Object 434239}: this object is classified as a merger because multiple components are present in both the [CII] map and in the UV-optical photometry. However, the \lya emission originates from the same region where the bulk of the [CII] emission is detected.
    
    \item {\bf Object 488399}: this object is classified as dispersion dominated, and the [CII] emission is much more extended than the optical light. This object is selected as a narrow-band emitter, with a very weak UV continuum, and therefore it is not detected in the optical images. For this object, we re-extracted the [CII] spectrum from a smaller region, coincident with the $K_s$ emission, but the position of the [CII] velocity peak did not change.
    
    \item {\bf Object 493583}: this object is classified as a merger mainly because two dynamical components are present in the position-velocity diagram, separated in velocity by $\sim$ 200 km s$^{-1}$ (see Le F\`evre et al. 2020). The \lya-UV-optical rest-frame and [CII] emissions are all spatially coincident.
    
    \item {\bf Object 494057}: this object is classified as a rotator, and the [CII] emission is much more extended than the \lya-optical emission. We re-extracted the [CII] spectrum from a smaller region, coincident with the \lya-optical emission, but the position of the [CII] velocity peak was not affected. There are no signs of multiple components from the images or the spectra.
    
    \item {\bf Object 494763}: this object is classified as dispersion dominated. The \lya-UV-optical rest-frame and [CII] emissions are all spatially coincident. The \lya spectral profile is double peaked, as it often is in models of homogeneous expanding shells (Verhamme et al. 2008). 
    
    \item {\bf Object 510660}: the [CII] emission for this object is very faint, and therefore it was not possible to obtain a morpho-kinematic classification. The [CII] spectrum is also quite noisy. The \lya-optical rest-frame emission is coincident with the peak of the [CII] emission. There are no signs of multiple components from the images or the spectra.
    
    \item {\bf Object 519281}: this object is classified as a merger because a faint second component is present in the position velocity diagram (Le F\'evre et al. 2020) and because the [CII] map shows a fainter companion southwest of the main emission (at the edge of the spectroscopic slit). We confirmed that these secondary peaks do not affect the position of the [CII] velocity peak. The \lya-optical rest-frame light is well aligned with the main [CII] emission.
    
    \item {\bf Object 536534}: this object is classified as a merger because the [CII] emission is resolved into three to four  clumps. The \lya-optical emission is spatially coincident with the brightest [CII] clump: we re-extracted the [CII] spectrum from this region alone, and we verifired that the [CII] velocity peak does not move (green line in the small inset).
    
    \item {\bf Object 539609}: this object is classified as a rotator. The \lya-optical emission is well aligned with the [CII] emission, and there is no evidence of multiple components. The spectral \lya-[CII] offset is also very small.
    
    \item {\bf Object 627393}: this object is classified as a merger because two distinct components are present in the position velocity diagram (Le F\'evre et al. 2020). However, these components are separated by $\sim$200 km s$^{-1}$ in velocity and by less than 0.5" in space. The \lya and [CII] spectra are extracted from the same physical region.
    
    \item {\bf Object 628063}: the [CII] emission for this object is quite faint, and no morpho-kinematic classification was performed. There is a spatial offset of $\sim$0.5" between the \lya-optical emission and the [CII] emission, but there is no evidence of multiple components in the images or spectra.
    
    \item {\bf Object 630594}: this object is classified as dispersion dominated, and the \lya emission is possibly slightly offset by $\sim$ 0.5" from the optical and the [CII] emissions.
    
    \item {\bf Object 665509}: this object is classified as a merger because a secondary [CII] peak lies next to the main peak. The \lya-optical rest-frame emission is coincident with the brightest [CII] clump. We re-extracted the [CII] spectrum for this bright clump alone, and the position of the [CII] velocity peak is not affected.
    
    \item {\bf Object 665626}: this object is too faint in [CII] for us to perform a morpho-kinematic classification. Its UV continuum is very faint, and therefore it is not detected in the optical band containing the \lya line. However, the optical rest-frame light, traced by theUltraVISTA $K_s$ image, is coincident with the peak of the [CII] emission. There is no evidence of multiple components.
    
    \item {\bf Object 680104}: this object is too faint in [CII] for us to perform a morpho-kinematic classification. The \lya-optical rest-frame emission is possibly offset by $\sim$0.5" from the peak of the [CII] emission, but there is no evidence of multiple components.
    
    \item {\bf Object 709575}: this object is classified as a rotator. The \lya-optical rest-frame light is well aligned with the [CII] emission, and there is no sign of multiple components.
    
    \item {\bf Object 733875}: this object is classified as dispersion dominated. The \lya-optical rest-frame light is well aligned with the [CII] emission, and there is no sign of multiple components.
    
    \item {\bf Object 742174}:  this object is too faint in [CII] for us to perform a morpho-kinematic classification. The \lya and optical emissions are offset by $\sim$0.7" from the peak of the [CII] emission. There is no evidence of multiple components.
    
    \item {\bf Object 773857}: this object is classified as a merger because a faint [CII] clump is detected alongside the main clump. The UV rest-frame continuum of this object, which is selected as a narrow-band emitter, is very faint, therefore it is not detected in the optical image. However, the optical rest-frame light originates from a region close to the peak of the [CII] emission. We verified that the secondary [CII] clump does not affect the position of the [CII] velocity peak.
    
    \item {\bf Object 803480}: this object is classified as dispersion dominated. The \lya-optical rest-frame light is well aligned with the peak of the [CII] emission, and the velocity offset \lya-[CII] is also very small. There is no evidence of multiple components.
    
    \item {\bf Object 814483}: this object is classified as a merger because the [CII] emission has many components. The UV and optical rest-frame light is aligned with only the brightest [CII] emitting region: we verified that the position of the [CII] velocity peak is not affected by the faintest clumps.
    
    \item {\bf Object 848185}: this object is classified as dispersion dominated. The [CII] emission is spatially extended: in the same region in which the [CII] is detected, several clumps are revealed in the UV rest-frame by the HST, but these clumps are not deblended in the optical rest-frame or in the spectroscopic data. The \lya and [CII] spectra are extracted from the same region.
    
    \item {\bf Object 859732}: this object is classified as a merger because multiple [CII] clumps are present. However, only the brightest [CII] clump lies in the spectroscopic slit where the \lya line has been extracted. The secondary [CII] clumps do not affect the position of the [CII] velocity peak.
    
    \item {\bf Object 873321}: this object is classified as a merger because the [CII] emission arises from two bright clumps. The \lya-optical emission is coincident with only one of the two clumps: in this case, we also re-extracted a [CII] spectrum from a smaller region, coincident with the region emitting the \lya-optical light, and we find that the position of the [CII] velocity peak does not change.
    
    \item {\bf Object 873756}: this object is classified as a merger because of multiple components in the ancillary optical Subaru images (and partly because of fainter [CII] structures around the main one). In particular, there is a bright UV-optical emitter southwest of the main [CII] emitter, just outside of the spectroscopic slit. Is important to stress that this companion is not responsible for the \lya emission that is extracted from within the slit, right on top of the main [CII] peak. We re-extracted a [CII] spectrum from a smaller region around the optical rest-frame emission, and we find that the [CII] velocity peak is not affected.
    
    \item {\bf Object 880016}: this object is classified as dispersion dominated. The \lya-optical emission is coincident with the peak of the [CII] emission.
    
    \item {\bf Object 881725}: this object is classified as a rotator, and the \lya-optical emission is coincident with the peak of the [CII] emission. The \lya spectrum is quite noisy and double-peaked.

\end{itemize}

We can draw a few conclusions from this analysis. First, in all cases that were not classified as mergers, we are quite sure that the [CII] emission, \lya, and the continuum arise from the same galaxy, although in many cases, a small offset is present, typically 0.5" at most. This is expected because the \lya emission is often found to be offset from the UV continuum by similar amounts (Leclercq et al. 2017; Hoag et al. 2019). These spatial offsets might imply that the ISM medium is inhomogeneous and clumpy, and \lya photons can more easily escape from one side of the galaxy than from the other.

For about half of the objects that are classified as mergers, it is evident that they are composed of two subgroups: cases in which multiple components are resolved both in ALMA [CII] images and UV-optical rest-frames (e.g., 417567, 536534, 665509, 859732, 873321, etc.), and cases in which the multiple components are unresolved in [CII] (e.g., 308643, 372292, 422677, 5101244930, etc.). For mergers in the first category, we stress that the \lya line is always extracted from a region coincident with the brightest [CII] clump, and we showed that the position of the [CII] velocity peak does not change, regardless of whether the [CII] spectrum is extracted from a small region around the region from which \lya is extracted. The analysis presented in this paper therefore applies to the {\it \textup{main component}} of the merger. Mergers in the other category are all classified as such because multiple components were identified in the position-velocity diagram, although these are not resolved in the [CII] map. Their [CII] emission is not very extended ($\sim$1"-1.5" across, corresponding to $\sim$6-9 kpc at $z\sim5$), and the different components are typically separated in velocity by $\sim$200 km s$^{-1}$. They can therefore be considered late-stage mergers, very close to coalescence. The \lya line is extracted from the same region that emits [CII]. For them, the \lya-[CII] and ISM-[CII] offsets are relative to the barycenter of the system, which is traced by the [CII] peak velocity, and the analysis presented in this paper applies to their {\it \textup{integrated}} extent.

  \renewcommand{\thefigure}{A.\arabic{figure} (Cont.)}
\addtocounter{figure}{-1} 
   \begin{figure*}
\centering
   \includegraphics[trim={0 1.5cm 0 1cm}, clip, width=0.96\textwidth ]{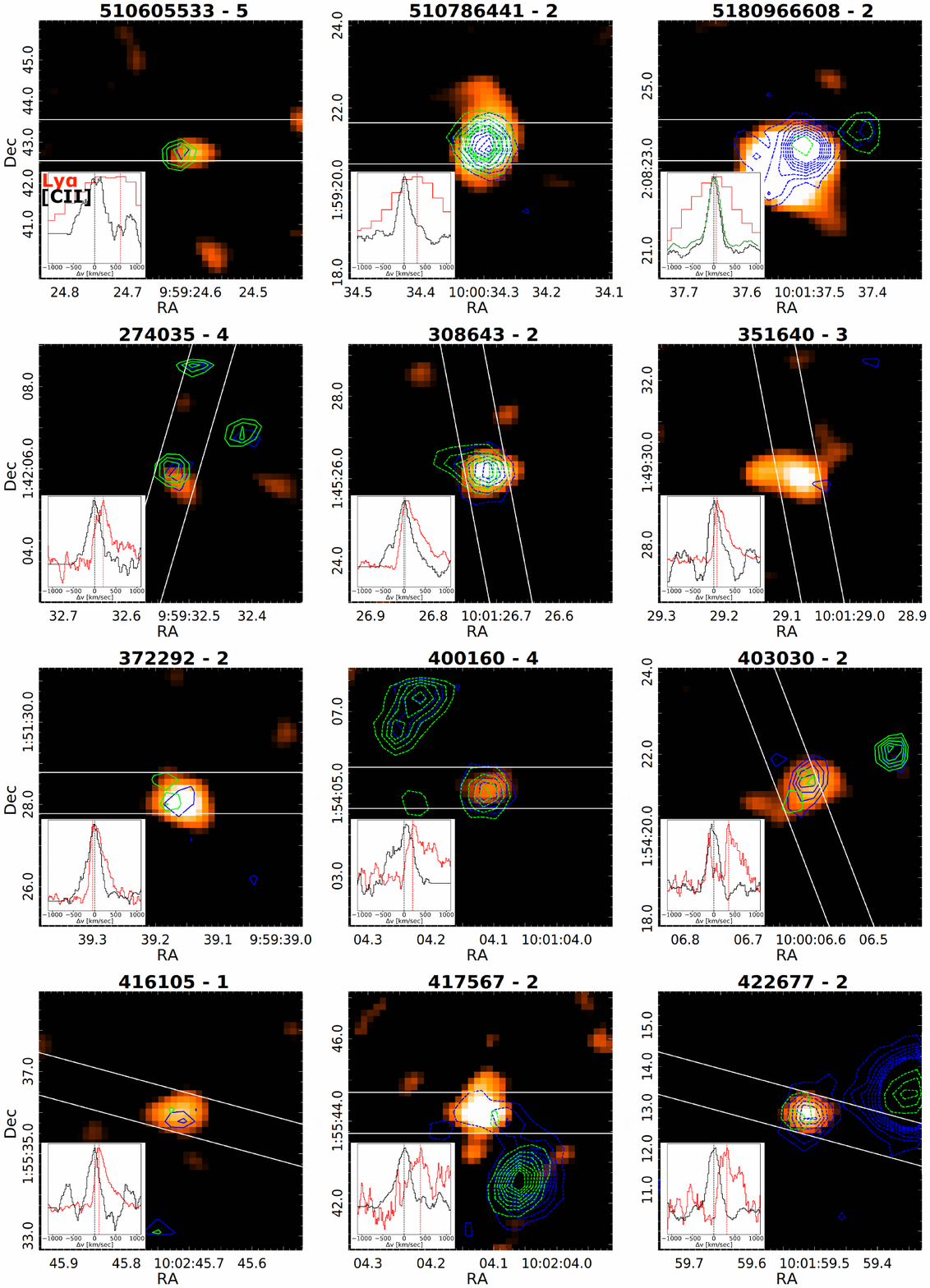}
    \caption{[CII flux maps with...]}
      \label{Fig:cutouts2}
   \end{figure*}
  \renewcommand{\thefigure}{A.\arabic{figure} (Cont.)}
\addtocounter{figure}{-1}    
   
   \begin{figure*}
   \centering
   \includegraphics[trim={0 1.5cm 0 1cm}, clip, width=0.96\textwidth ]{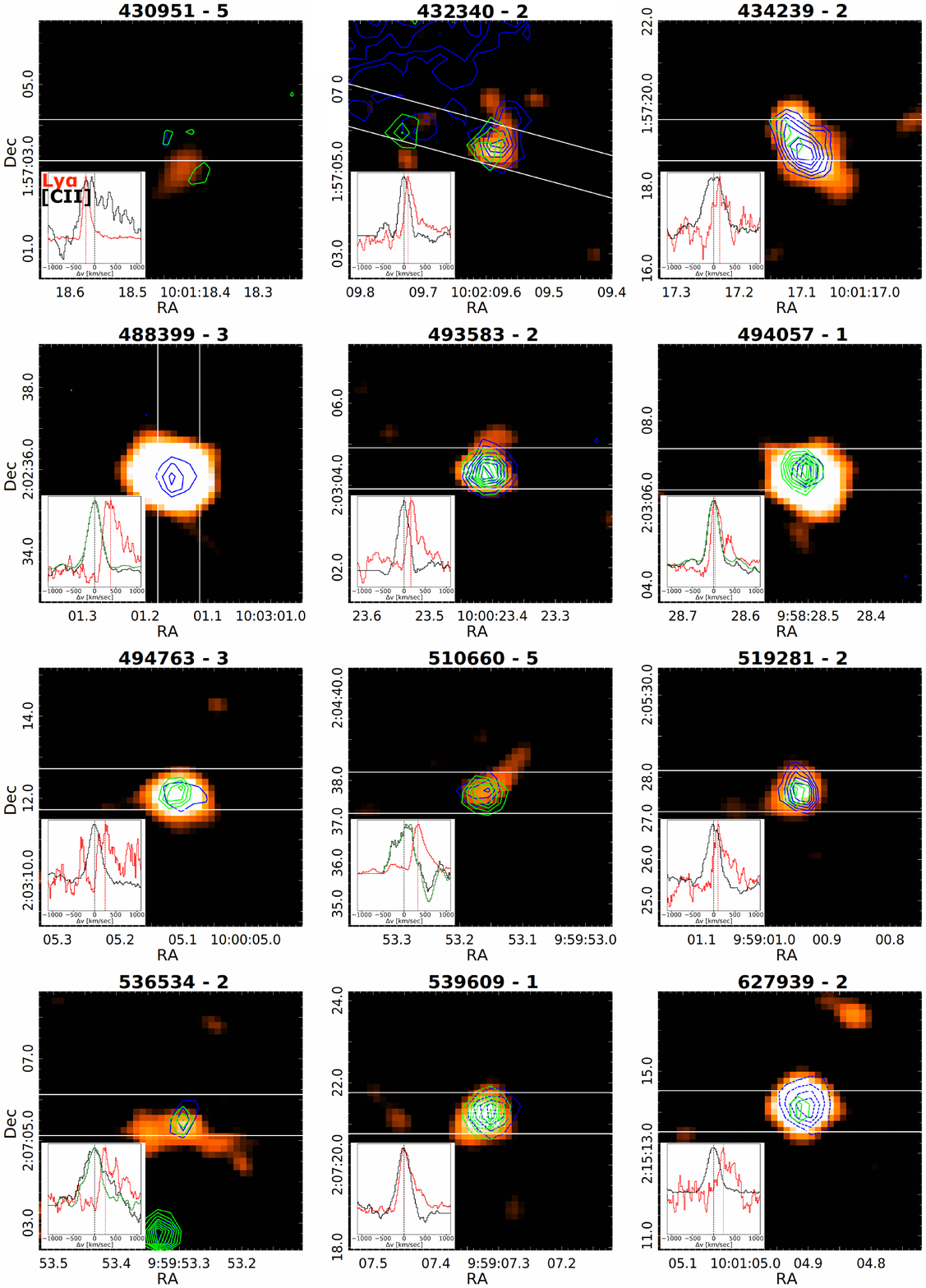}
   \caption{[CII flux maps with...]}
      \label{Fig:cutouts3}
   \end{figure*}
  \renewcommand{\thefigure}{A.\arabic{figure} (Cont.)}
\addtocounter{figure}{-1}    

   \begin{figure*}
   \centering
   \includegraphics[trim={0 1.5cm 0 1cm}, clip, width=0.96\textwidth ]{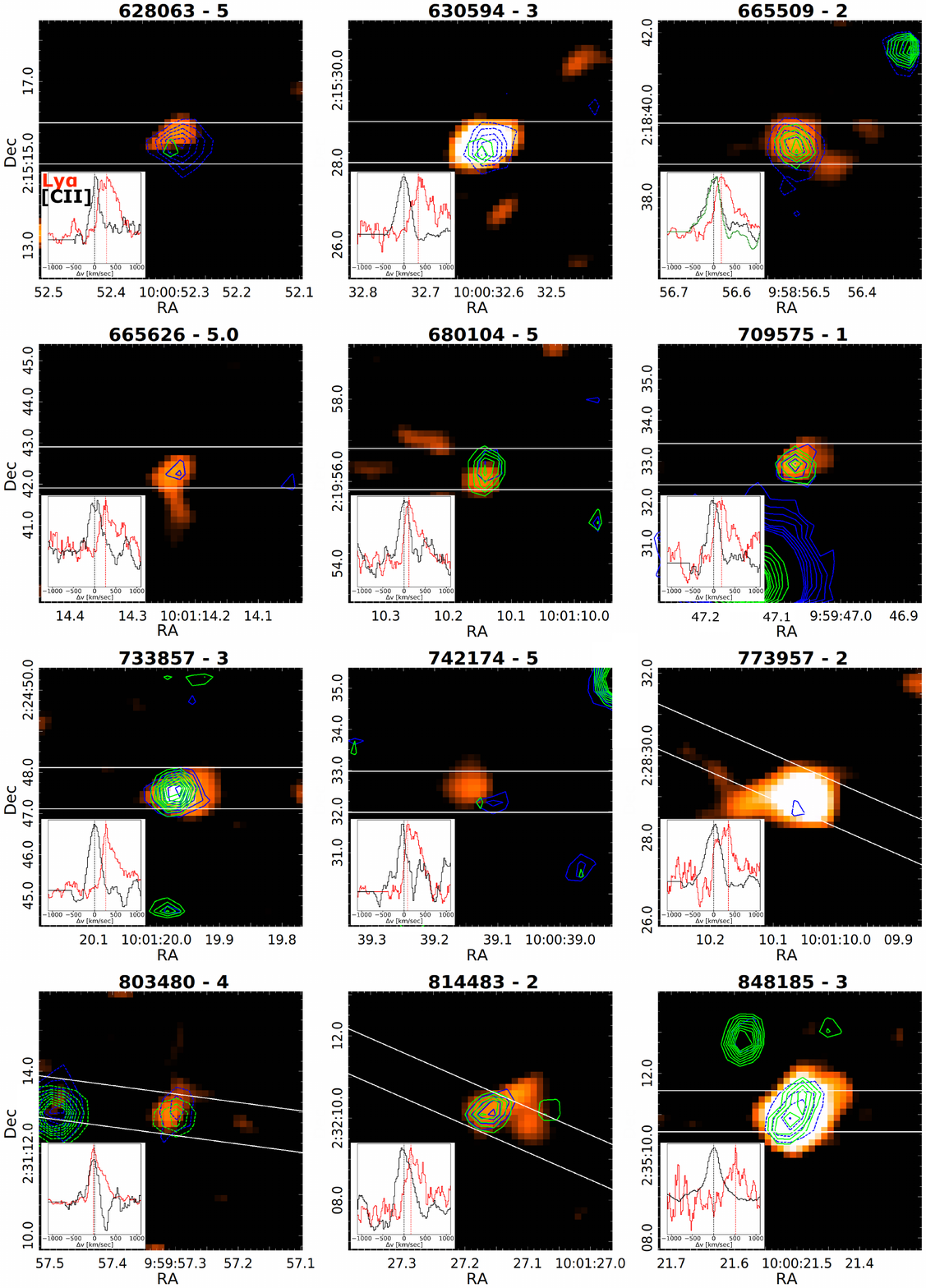}
   \caption{[CII flux maps with...]}
      \label{Fig:cutouts4}
   \end{figure*}

   \renewcommand{\thefigure}{A.\arabic{figure} (Cont.)}
\addtocounter{figure}{-1}    
  
      \begin{figure*}
   \centering
   \includegraphics[trim={0 15cm 0 1cm}, clip, width=0.96\textwidth ]{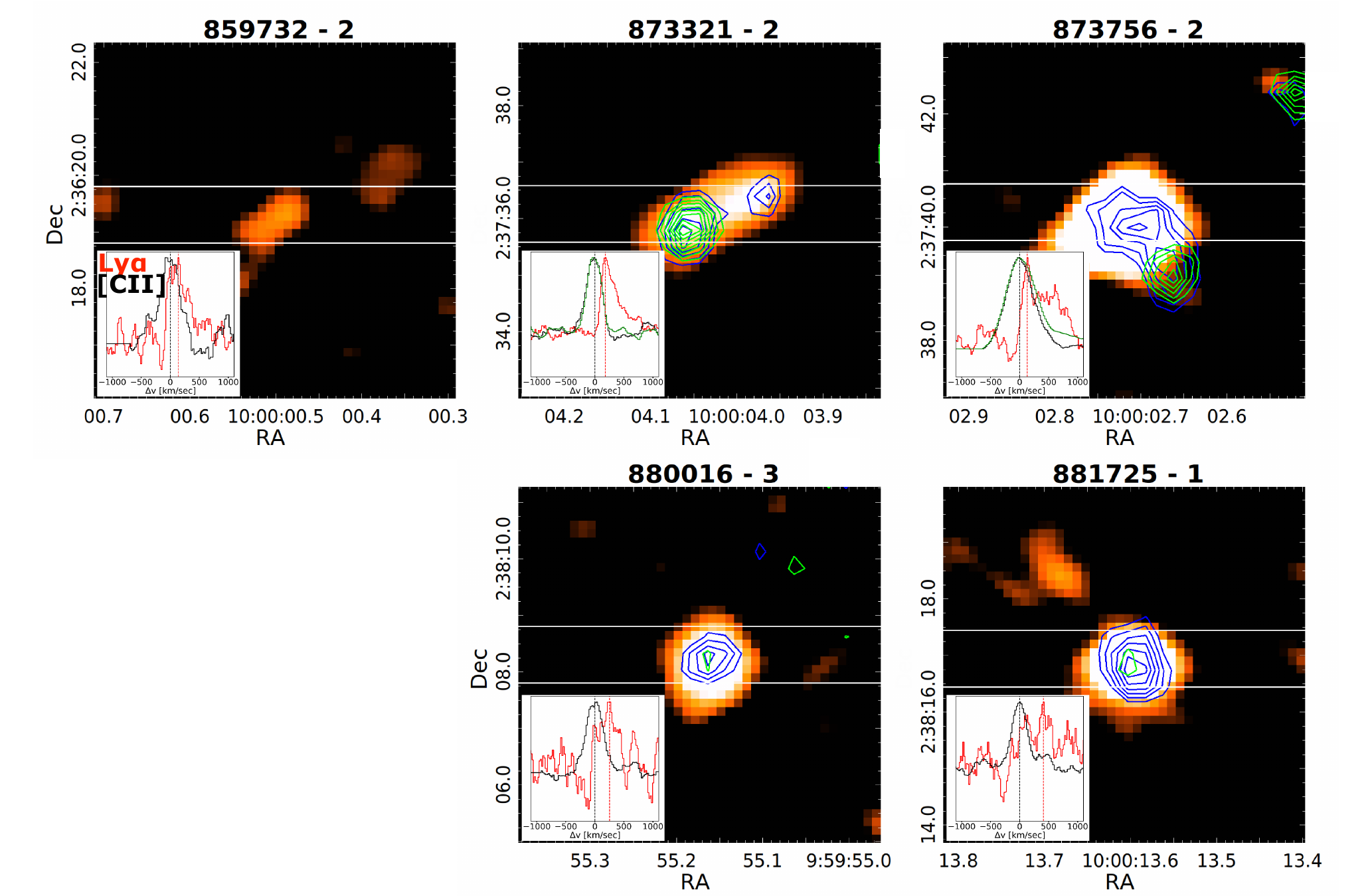}
   \caption{[CII flux maps with...]}
      \label{Fig:cutouts5}
   \end{figure*}

\end{appendices}

%
%

\end{document}